\documentclass[12pt,preprint]{aastex}

\newcommand{\be}{\begin{equation}}    
\newcommand{\ee}{\end{equation}}
\newcommand{\ba}{\begin{eqnarray}}
\newcommand{\ea}{\end{eqnarray}}

\catcode`@=11
\newcommand\eqalign[1]{\null\,\vcenter{\openup\jot\m@th
    \ialign{\strut\hfil$\displaystyle{##}$&$\displaystyle{{}##}$\hfil
        \crcr#1\crcr}}\,}

\slugcomment{Submitted to Ap.J., May/7/2007.}

\begin{document}


\title{On the Orbit Structure\\ of\\
    the Logarithmic Potential}

\author{C. Belmonte}
\affil{Physics Department, University of Rome ``La Sapienza", P.le A. Moro, 2 -- I00185 Rome}
\email{belmonte@roma1.infn.it}

\author{D. Boccaletti}
\affil{Mathematics Department, University of Rome ``La Sapienza", P.le A. Moro, 2 -- I00185 Rome}
\email{boccaletti@uniroma1.it}

\and

\author{G. Pucacco\altaffilmark{1}}
\affil{Physics Department, University of Rome ``Tor Vergata", Via della Ricerca Scientifica, 1 -- I00133 Rome}
\email{pucacco@roma2.infn.it}

\altaffiltext{1}{INFN - Sez. Roma Tor Vergata}

\begin{abstract}
We investigate the dynamics in the logarithmic galactic potential with an analytical approach. The phase-space structure of the real system is approximated with resonant detuned normal forms constructed with the method based on the Lie transform. Attention is focused on the properties of the axial periodic orbits and of low order `boxlets' that play an important role in galactic models. Using energy and ellipticity as parameters, we find analytical expressions of several useful indicators, such as stability-instability thresholds, bifurcations and phase-space fractions of some orbit families and compare them with numerical results available in the literature.

\end{abstract}



\section{Introduction}
To determine salient features of the orbital structure of non-integrable potentials is a fundamental topic in galactic dynamics. Knowledge of existence and stability of periodic orbits of low commensurability is for example very important to clarify the issue of triaxiality in elliptical galaxies and a general understanding of the phase-space structure is very useful in applications of self-consistent models. The numerical approach is usually preferred due to the availability of reliable algorithms and powerful machines (Schwarzschild, 1979; Merritt \& Valluri, 1996; Papaphilippou \& Laskar, 1996 and 1998). However, in several circumstances, is useful to have some simple analytic clues concerning the relation between the form of the gravitational potential and the main families of orbits supported, as, for example, in the studies by Zhao et al. (1999). 

Although generically non-integrable, realistic galactic potentials show several properties of a regular behavior over wide energy ranges, with invariant surfaces (``tori'') surrounding periodic orbits. Among these features, information about the stability properties of the main periodic orbits is of paramount importance, since the bulk of density distribution is shaped by the stars in regular phase-space domains around stable periodic orbits (Binney \& Tremaine, 1987). In particular, for triaxial ellipsoids, periodic orbits along symmetry axes play a special role. An enormous effort has therefore been devoted to investigate families and bifurcations of periodic orbits, starting with the study of models based on perturbed oscillators (Contopoulos, 1970) and gradually exploring more realistic galactic potentials with numerical (Miralda-Escud\'e \& Schwarzschild, 1989; Fridman \& Merritt, 1997) and semi--analytical (de Zeeuw \& Merritt, 1983; Scuflaire, 1995) approaches. A related important problem is that of torus construction in the regular domains (Binney  \& Spergel, 1982; Gerhard \& Saha, 1991; Kaasalainen \& Binney, 1994a,b).

Techniques based on the various versions of perturbation theory have been applied to several examples and with different degrees of approximation (for a review, see, e.g., Contopoulos, 2002). One of the most powerful analytic tools is the normal form approximation of a non integrable system. Although the normal form approach is quite widespread in galactic dynamics, its use in studying stability of periodic orbits has not been as systematic as the theory could allow (Sanders \& Verhulst, 1985). Aim of the present paper is to apply the Lie transform normalization method (Hori, 1966; Deprit, 1969; Dragt \& Finn, 1976; Finn, 1984; Koseleff, 1994) to approximate the dynamics of the Binney logarithmic potential (Binney, 1981). We compare the findings with that of Miralda-Escud\'e \& Schwarzschild (1989), who employ purely numerical techniques to implement the Floquet method and with that of Scuflaire (1995), who studies the stability of axial orbits by solving the Hill-like perturbation equation with the Lindstedt--Poincar\'e approach. Another example that is briefly mentioned is provided by the galactic Schwarzschild (1979) potential with a comparison with the results of de Zeeuw \& Merritt (1983). These authors based their approach on the averaging procedure of normalization: it is therefore interesting a comparison with that method as well. 

To study the structure of phase space with a truncated normal form one may proceed in essentially two ways: the most general and exhaustive is that of determining the explicit form of periodic orbits and invariant tori that give the `skeleton' of the phase-space of the system. A less general but easier approach in 2-degrees of freedom (DoF) systems, is that of determining fixed points and invariant curves on a surface of section. This is constructed with the aid of the approximate integral of motion provided by the normalization. Even in the case of linear orbital stability of the main periodic orbits, the first method is in general quite cumbersome and can be applied when the procedure of reduction to a single degree of freedom Hamiltonian system and the use of action--angle variables lead to a reasonably simple system of equations. This reduction is always possible with 2 DoF, both in the resonant and non-resonant cases (Gustavson, 1966), whereas it is in general no more possible in resonant 3-DoF systems. The second method is clearly less general but relies on straightforward geometric arguments related to the Hessian of a polynomial in its critical points and is, at least in principle, quite easy to implement. In this work we are going to apply both methods to perform the comparison mentioned in the paragraph above. There is another motivation to linger on both visions of the approximate dynamics: the normal form Hamiltonian is, in many respect, the ``simplest'' integrable relative of the non-integrable original one. However, this simplicity is attained at the price of a set of coordinate transformations that deform the phase-space variables; if one is interested in specific local properties or wants to make comparison with numerical simulations, the inversion to the original variables becomes mandatory. Therefore we want to deepen the relation between the `clean' but deformed dynamics provided by the normalizing variables and the dynamics in the physical variables `dirtied' by the inverse transformation.

Even in realistic potentials, the periodic orbits along the axes of symmetry (axial orbits) are easily identified both as normal modes of the reduced system and as ``central'' fixed points on the surfaces of section. Therefore, we will limit the detailed evaluation of the stability characteristics in the parameter space to these axial orbits. However, both procedures we have followed are quite general and can be directly applied to all periodic orbits of sufficiently low commensurability. From the results obtained, we can state that the predictive power of the normal form ranges well outside the neighborhood in which the expansion of the original Hamiltonian is performed. It is rather related to the extent of the asymptotic convergence radius of the approximate integrals of motion, namely the extent at which the remainder of the series truncated at a given order is minimal (see, e.g., Efthymiopoulos, Giorgilli \& Contopoulos, 2004). The results suggest that there is still room for improvement up to a certain {\it optimal} order of truncation. Therefore, at least in the cases we have examined, the regular dynamics of a non-integrable system can be reproduced with very high accuracy. However, in concrete applications, the validity of the prediction has to be corroborated with an independent evaluation of the best suited resonant normal form for the problem at hand, always bearing in mind that, after the onset of chaos, these methods become ineffective.

The plan of the paper is as follows: in Section 2 we recall the procedure of normalization as applied to reflection symmetry potentials and state a criterion for the choice of the best suited resonant normal form; in Sections 3 and 4 we study the 1:1 and 1:2 resonances respectively; in Section 5 we compare our analytical results with those available in the literature; in Section 6 we sketch the conclusions.

\section{Analytical set-up}

The normal form approach to investigate Hamiltonian systems has a wide range of applications in the astrophysical context. However, the method based on Lie transforms, which has several computational virtues with respect to the original methods (Gustavson, 1966; Contopoulos, 2002), has seldom been employed in galactic dynamics. Notable exceptions are the papers by Gerhard \& Saha (1991) and Yanguas (2001) concerning perturbed isochrone models.

In this section we recall the most relevant mathematical tools, with the aim of laying down the notations and the main formulas. For a detailed account of the method we refer to tetxbooks (Sanders \& Verhulst, 1985; Meyer \& Hall, 1992; Boccaletti \& Pucacco, 1999).

\subsection{General}

Suppose the system under investigation is given by a Hamiltonian
\be
   H({\bf{p,q}})= \frac{1}{2}(p_x^2+p_y^2) + V(x^2,y^2),
 \ee
with $V$ a smooth potential with an absolute minimum in the origin and reflection symmetry with respect to both axes. We expand the potential in the form
\be\label{potexp}
   V( x,y;\varepsilon)= 
   \sum_{k=0}^{\infty} \varepsilon^{k} V_{k}(x,y)
\ee
and look for a new Hamiltonian given by 
\begin{equation}\label{HK}
     K({\bf{P,Q}};\varepsilon)=\sum_{n=0}^{\infty}\varepsilon^n K_n ({\bf{P,Q}})=M_g^{-1}H({\bf{p,q}};\varepsilon),
  \end{equation}
  where $\bf{P,Q}$ result from the canonical transformation
  \be\label{TNFD}
  ({\bf{P,Q}}) = M_g ({\bf{p,q}}).\ee
In these and subsequent formulas we adopt the convention of labeling the first term in the expansion with the index zero: in general, the `zero order' terms are quadratic homogeneous polynomials and terms of {\it order} {\it n} are polynomials of degree $n+2$. The linear differential operator $M_g$ is defined by
\begin{equation}\label{eqn:OperD-F}
    M_g\equiv e^{-\varepsilon L_{g_1}}e^{-\varepsilon^2 L_{g_2}}\cdots e^{-\varepsilon^n L_{g_n}}\cdots = \sum_{n=0}^{\infty}\varepsilon^n M_n.
\end{equation}
The functions $g_n$ are the coefficients in the expansion of the generating function of the canonical transformation and the linear differential operator $L_{g}$ is defined through the Poisson bracket
\begin{equation} \label{eq:OdL}
    L_g f \equiv \{g,f\} \equiv 
    {{\partial{g}} \over {\partial {x}}}
    {{\partial{f}} \over {\partial {p_x}}} +
    {{\partial{g}} \over {\partial {y}}}
    {{\partial{f}} \over {\partial {p_y}}}- 
    {{\partial{g}} \over {\partial {p_x}}}
    {{\partial{f}} \over {\partial {x}}}-
    {{\partial{g}} \over {\partial {p_y}}}
    {{\partial{f}} \over {\partial {y}}}.
\end{equation}
The exponentials in the definition of $M_g$ are intended as the {\it formal} sum of a power series so that it gives rise to a near-identity coordinate transformation known as {\it Lie series}. Therefore, in practice, in the algorithm implemented to compute the $K_{n}$ appearing in (\ref{HK}), what are really used are the differential operators
\begin{equation}\label{eqn:OperMn}
    M_n =  \sum_{m_{1} + 2 m_{2} + \dots n m_{n} = n} \frac{L_{g_1}^{m_{1}} L_{g_2}^{m_{2}} \cdots L_{g_n}^{m_{n}}}{m_{1}! m_{2}!\cdots m_{n}!}.
\end{equation}
By expanding also the right hand side of (\ref{HK}) in power series of $\varepsilon$ and equating the coefficients of the same order, one has the recursive set of linear partial differential equations
\be\label{EHK} \begin{array}{ll}
    &K_0=H_0 = \frac{1}{2}(p_x^2+p_y^2) + V_{0}\, ,\\ \\
    &K_1=V_1+M_{1}H_{0} \, ,\\ \\
   &\quad \;\; \vdots \\ \\
    &K_n= V_n +M_nH_0 +\sum_{m=1}^{n-1}M_{n-m}V_m \, ,\\ \\
   &\quad \;\; \vdots 
\end{array}\ee
It can be seen that each equation in the chain (\ref{EHK}) depends only on quantities found in the previous line. `Solving' the equation at the $n$-th step consists of a twofold task: to find $K_{n}$ {\it and} $g_n$. The unperturbed part of the Hamiltonian, $H_{0}$, determines the specific form of the transformation. In fact, the new Hamiltonian $K$ is said to be {\it in normal form} if
\be\label{NFD}
\{H_0,K\}=0.
\ee
This condition is used at each step of the procedure to determine the functions $g_n$ in order to eliminate as much as possible terms in the new Hamiltonian. The only terms of which $K$ is made of are those staying in the kernel of the operator $L_{H_{0}}$ associated to $H_{0}$ through definition (\ref{eq:OdL}). The procedure is stopped at some ``optimal'' order $N$ and therefore in all ensuing discussion we refer to a ``truncated'' normal form.
$H_0$ must be considered a function of the new coordinates at each step in the process: according to (\ref{NFD}), it is therefore an integral of the motion for the new Hamiltonian $K$. The function 
\be\label{integral}
{\cal I} = K - H_{0} \ee
can be therefore used as a second integral of motion conveying approximate information on the dynamics of the original system. 

The normalizing transformation (\ref{TNFD}) leads to new coordinates which are continuous differentiable {\it deformations} of the original ones. 
For practical applications (for example to compare results with numerical computations) it is useful to express approximating functions in the original physical coordinates. Inverting the coordinate transformation, the new integral of motion can be expressed in terms of the original variables. Denoting it as the power series
\be\label{PHI}
 I = \sum_{n=0}^{\infty}\varepsilon^n I_n,\ee
its terms can be recovered by means of 
\be
I_n=V_n-K_n +\sum_{m=1}^{n-1}M_{n-m}\big[V_m-I_m\big] \, , \qquad n\geq 1\, ,
\label{eqn:In}\ee
that is obtained from (\ref{integral}) by exploiting the nice properties of the Lie transform with respect to inversions (Boccaletti \& Pucacco, 1999). 

It is important to remind in which respect the normal form $K$ and the integral $I$ provide approximations to the dynamics of the original system. The normal form is truncated at step $N$: this means that in the new Hamiltonian a ``remainder'' of order ${\rm O}(\varepsilon^{N+1})$ is neglected 
and the two functions $K$ and ${\cal I}$ provide an (exactly) integrable system whose dynamics, in the new coordinates, is ${\rm O}(\varepsilon^N)$ close to that of the original system. On the other hand, inverting the canonical normalizing transformation, the power series (\ref{PHI}) is an approximate integral of motion of the original system (in the original coordinates) since
\be
\{I,H\}={\rm O}(\varepsilon^{N+1}).
\ee
That is, $I$ fails to commute with the original Hamiltonian for terms of higher order. We see that in both cases the error made with the perturbative method is of the same amount since terms of the same order are neglected. However, it is not clear if, in the case of high orders and when a direct comparison is possible, both approaches lead exactly to the same predictions. Since in various applications one or the other approaches have pros and cons, one of the aim of the present work is just to examine and reconcile possible discrepancies.

We remark that in all subsequent applications involving series expansions, the role of the perturbation parameter can also be played by the size of the neighbourhood of the origin where the Hamiltonian is considered. Therefore the powers of the parameter $\varepsilon$ are left in all expansion formulas just to indicate their order and are treated as unity in the computations. 

\subsection{Expanding ``cored'' galactic potentials}

The model potential we will consider is the Binney logarithmic potential (Binney \& Tremaine, 1987). The logarithmic potential has played a fundamental role in the description of galactic models displaying, with a very simple analytical expression, many realistic features of elliptical galaxies, in particular employing its singular scale-free form (Miralda-Escud\'e \& Schwarzschild, 1989)
\be\label{SLP}
 V = \frac12 \log(x^2 + y^2 / q^2).\ee
 We cannot apply the standard normalization procedure as it is to a singular potential. In this way we are still not able to explore the phase-space structure of cuspy models that nowadays seem to be required by the observed properties of real galaxies (Merritt, 1999). However, selected classes of power law or other singular potentials can be put in a form suitable to the normalization algorithm by a proper coordinate transformation (Sridhar \& Touma, 1997). Moreover, it is important to remark that the class of ``weak'' cusps (Dehnen, 1993) displays many features of cored potentials as can be seen in the numerical exploration performed by Fridman \& Merritt (1997). Therefore, the analysis presented here, even if restricted to potentials that can be Taylor expanded around a homogeneous density core, is a useful starting point. Enforcing the reflection symmetries of galactic potentials, we assume that each term in the expansion (\ref{potexp}) can be written as a homogeneous polynomial of degree $2(k+1)$ of the form
\be\label{polexp}
V_{2k}( x,y) =  \frac{1}{2(k+1)} \sum_{j=0}^{k+1} a_{2(j,k-j+1)} x^{2j} y^{2(k-j+1)}, \quad k=0,1,2,\dots\ee 
 A Taylor expansion has a convergence radius that can be finite and small if compared with the region of interesting dynamics. One could conservatively deduce that the normal form is a good appoximation of the dynamics only {\it within} the convergence radius of the potential. However, things are not so simple, because what really matters is the convergence of the normal form itself. In fact it may happen that a potential with infinite convergence radius, for example a perturbed oscillator without escape, has a normal form with a finite convergence radius: usually, this is related to the break up of regular dynamics and the transition to stochasticity. Clearly, in this case there is no hope to get reliable information on the dynamics beyond the stochasticity threshold from the normal form. On the opposite side, there is the possibility that the system, even if non integrable, displays regular dynamics almost everywhere and in this case the normal form can be used even {\it outside} the convergence radius of the original potential.  
 The logarithmic potential offers the opportunity to explore all these issues (Kaasalainen \& Binney, 1994b).  

The cored logarithmic potential can be written as 
\be\label{LP}
 V = \frac12 \log(1 + x^2 + y^2 / q^2).\ee
The form written here is simplified by the choice of fixing the length scale (the ``core radius'' $R_{c}$) equal to one, but this is not a limitation due to the invariance in both the length scale and the energy scale. With these units, the energy $E$ may take any non negative value 
 \be 0 \le E < \infty. \ee
 The parameter giving the ``ellipticity'' of the figure ranges in the interval
 \be\label{rangeq}
 0.6 \le q \le 1. \ee
 Lower values of $q$ can in principle be considered but correspond to a non physical density distribution. Values greater than unity are included in the treatment by reversing the role of the coordinate axes, effectively extending the range until the upper limit $1/0.6 = 5/3$. The series expansion of the logarithmic potential is 
\be\label{ELP}
 V = \frac12 \sum_{k=0}^{\infty} \frac{(-1)^{k}}{k+1} \left(x^2 + \frac{y^2}{q^2} \right)^{k+1},
 \ee
so that the coefficients appearing in (\ref{polexp}) are
\be\label{alog}
a_{2(j,k-j+1)}={{k+1}\choose{j}} \frac{(-1)^{k}}{q^{2(k-j+1)}}.
\ee
Those of lowest order are
\ba
&& a_{(2,0)} = \omega_1^{2}=1,\;\; a_{(0,2)} = \omega_2^{2}=1/q^{2}, \label{alog1} \\
&& a_{(4,0)} = -1,\;\;
a_{(2,2)} = -2/q^2,\;\;
a_{(0,4)} = -1/q^4, \label{alog2} \\
&& a_{(6,0)} = 1,\;\;
a_{(4,2)} = 3/q^2,\;\;
a_{(2,4)} = 3/q^4,\;\;
a_{(0,6)} = 1/q^{6}. \label{alog3} 
\ea
 The natural setting in which one performs a low order normalization according to definition (\ref{NFD}) is therefore that of a perturbed quadratic Hamiltonian with a potential starting with
the harmonic term and `frequencies' given by (\ref{alog1}). For a generic value of $q$, the frequency ratio
\be
\frac{\omega_1}{\omega_2} \equiv q\ee
is a real number: the frequencies are rationally independent and the terms in the normal form, that is those polynomials in the canonical variables commuting with 
\be\label{Hzero}
H_{0} = \frac12 (p_x^2 + p_y^2 + \omega_1^2 x^2 + \omega_2^2 y^2),
\ee
consist only of functions of the partial energies in the harmonic potential. It is customary to refer to the normal form constructed in this case as a ``Birkhoff'' normal form (Birkhoff, 1927). The presence of terms with small denominators in the expansion, forbids in general its convergence. It is therefore more effective to work since the start with a {\it resonant normal form} (Sanders \& Verhulst, 1985), which is still non convergent, but has the advantage of avoiding the small divisors associated to a particular resonance. To catch the main features of the orbital structure, we therefore approximate the frequencies with a rational number plus a small ``detuning'' that we assume ${\rm O}(\varepsilon^2)$
\be\label{DET}
\frac{\omega_1}{\omega_2} = q = \frac{m}{n} + \varepsilon^2 \delta. \ee
We speak of a {\it detuned} ({\it m:n}) {\it resonance}, with $m+n$ the {\it order} of the resonance. The algorithm to perform a resonant normalization generalizes that of the Birkhoff normalization in its ability to identify additional terms to be included in the normal form, since they cannot be eliminated with the sequence of canonical transformations. To this aim, the system must be in a form suitable to apply the resonant normalization procedure: we rescale variables according to
\be
x \longrightarrow {\sqrt{\omega_1}} x,\quad  
y \longrightarrow {\sqrt{\omega_2}} y, \quad
p_x \longrightarrow \frac{p_x}{\sqrt{\omega_1}},\quad  
p_y \longrightarrow \frac{p_{y}}{\sqrt{\omega_2}},\ee
in order to put the Hamiltonian in the form
\be\label{detH}
{\cal H} = \frac12 [(m+ n \varepsilon^2 \delta) (p_x^2+x^2) + n (p_y^2+y^2)]+ \sum_{k=0}^{\infty}  
\frac{\varepsilon^k }{2(k+1)}
\sum_{j=0}^{k+1} b_{2(j,k-j+1)} x^{2j} y^{2(k-j+1)}\ee
where we have used the same notation for the rescaled variables and
\be \label{abc}
b_{2(j,k-j+1)}=  \frac{n a_{2(j,k-j+1)}}{\omega_1^{j} \omega_2^{k-j+2}} . \ee
Frequencies have been approximated as in (\ref{DET}) and the Hamiltonian is redefined according to
the rescaling
\be\label{newE}{\cal H} := \frac{n H}{\omega_2} = n q H.\ee
As before, the new Hamiltonian $K$ is said to be in normal form if
\be
\{{\cal H}_0,K\}=0
\ee
where, in agreement with (\ref{newE}), the unperturbed part of the Hamiltonian (\ref{Hzero}) has been replaced by 
\be
{\cal H}_0=\frac12 [m (p_x^2+x^2) + n (p_y^2+y^2)]\ee
and the procedure is now that of an ordinary resonant ``Birkhoff--Gustavson'' normalization (Gustavson, 1966; Moser, 1968) with two variants: the coordinate transformations are performed through the Lie series and the detuning quadratic term is treated as a term of higher order and put in the perturbation.

\subsection{Choice of the resonance}

The ``skeleton'' of the regular part of the phase-space of a non integrable system is framed by periodic orbits and invariant tori. In particular, knowing location and stability of periodic orbits allows one to gather a substantial piece of information concerning the whole dynamics. Moreover, determining the instability thresholds in terms of physical parameters (e.g., the energy), provides also clues on the extent at which stochasticity becomes important. The change in stability of a periodic orbit is connected to frequency ratios: commensurability of low orders between frequencies is the main trigger to stability-instability transition and the interaction of resonances provides chaos. In harmonic oscillators, frequency ratios are fixed: if their ratio is non-rational there is no resonance. For example, the quadratic part of the logarithmic potential is given by coefficients (\ref{alog1}); with $q$ in the range (\ref{rangeq}) there is no reason to assume it to be a rational number. However, coupling between the degrees of freedom due to the perturbation causes the frequency ratios to change. The system {\it passes through} resonances of order given by the integer ratios closest to the ratio of the unperturbed frequencies. Crossing the resonance, a stability change may occur and, in general, a new sequence of periodic orbits bifurcates (Merritt, 1999; Contopoulos, 2002). Given an arbitrary pair of unperturbed frequencies, it could seem better to approximate their ratio as close as possible with integers. However there is an argument on which a more effective choice can be based. A typical situation is that in which a family of periodic orbits becomes unstable when a low order resonance occurs between its fundamental frequency and that of a normal perturbation: the simplest case is given by an axial orbit that, depending on the specific form of the potential, can be unstable through bifurcation of loop orbits (1:1 resonance), ``banana'' orbits (1:2 resonance), ``fish'' orbits (2:3 resonance), etc. Therefore, a detuned low-order resonant normal form can be quite accurate in describing the corresponding bifurcations. The strategy can be either a systematic exploration of the hierarchy of resonances or a specific choice guided by some independent argument (energy range, ellipticity range, etc.). Clearly, since the frequency ratio (\ref{DET}) is given by the parameter $q$ that determines the shapes of the isopotentials, fixing a specific value of $q$ in the range (\ref {rangeq}) also indicates which resonance is expected to be closer.
 
 It must be emphasized that the structure of a resonant normal form is also affected by the symmetries of the original system. The normal form must preserve these symmetries and this in general also leads to a criterion for truncation. In the present instance of a double reflection symmetry, given a resonance ratio $m/n$, the normal form must contain at least terms of degree $2\times(m+n)$ (see, e.g. Tuwankotta \& Verhulst, 2000). Therefore, the criterion we have adopted in this paper has been that of starting with the lowest order truncated normal form incorporating the symmetries of a typical galactic potential. A systematic investigation of the optimal order of truncation has recently been performed by Contopoulos et al. (2003) and Efthymiopoulos et al. (2004). Their results confirm the rapid decrease of the optimal order with the radius of the phase-space domain in which expansions are computed: we may conjecture that if we are interested in the global dynamics and accept a moderate level of accuracy, with this very conservative approach we can get reliable information up to the breakdown of the regular dynamics. Once particular orbits (or other interesting features) of the system are selected, more accurate details concerning the location, stability thresholds and so forth, can be derived with a higher order truncation. 



\section{1:1 symmetric resonance and second order normalization}

We proceed by exploiting the normal form of the system with Hamiltonian (\ref{detH}), potential (\ref{ELP}) and prescription (\ref{abc}), starting with the simplest possibility: a Lie transform normalization truncated to the first non-null term in the normal form. Recalling the symmetries of the potential represented by (\ref{alog1}), the first non trivial equation in the chain (\ref{EHK}) is
\be\label{EHK2}
K_2=V_2+M_{2}{\cal H}_{0},\ee
so that we actually truncate at order 2. Applying standard methods (Meyer \& Hall, 1992; Boccaletti \& Pucacco, 1999) gives the following expression of the second-order normal form (Belmonte et al. 2006)
\ba
&& K_{2}^{(1:1)} = \frac12 \delta (P_X^2+ X^2) - \frac{3}{32} \left(q({P_X}^2 + {X}^2)^2 + \frac{1}{q}({P_Y}^2 + {Y}^2)^2 \right) \nonumber \\
  &&           - \frac{1}{16}   \left( {P_Y}^2 (3 {P_X}^2 + {X}^2) + {Y}^2  ({P_X}^2 + 3                   {X}^2) + 4 P_X P_Y X Y \right) , \label{K211}\\
  && K_{2}^{(m:n)} = \frac{n}{2} \delta (P_X^2+X^2) + k_{1} \left(q({P_X}^2 + {X}^2)^2 + \frac{1}{q}({P_Y}^2 + {Y}^2)^2\right) \nonumber \\
&&  \quad \quad \quad + k_{2}  ({P_X}^2 + {X}^2) ({P_Y}^2 + {Y}^2) ,  
\label{K2mn}  
\ea
where the values of the coefficients in the expansion of the potential come from (\ref{alog1}) and (\ref{alog2}) and $k_{1}, k_{2}$ are rational numbers dependent on $m$ and $n$. Eq.(\ref{DET}) has been used 
and the canonical variables $\bf{P,Q}$ are as in Eq.(\ref{HK}). We see that this case behaves in the same way as in the first order averaging approach (de Zeeuw \& Merritt, 1983; Verhulst, 1996): the 1:1 resonance {\it or all other resonances}. This remark is another clue to the strategy delineated in subsection 2.3, that is to limit the global analysis to a normal form truncated to the first term incorporating a meaningful resonance. Therefore, for a preliminary investigation of periodic orbits with the lowest commensurability, that in the present instance is
\be\label{DET11}
\frac{\omega_1}{\omega_2} = 1+ \varepsilon^2 \delta, \ee
we can just exploit a $1:1$ normal form truncated at $K_{2}^{(1:1)}$. 

In the following subsections we work out in detail this analysis to take it also as a record for the cases with higher order resonances. In fact, in these cases, the procedure goes along the same lines, but reporting computations in details is not so easy due to huge expressions. In Sect. 5.1 we will get more accurate numerical results going to order 6. A detailed study of the orbits in the 1:1 resonance case (including periodic and non periodic orbits) has been provided by Contopoulos (1965).

\subsection{Orbit structure of the 1:1 resonance}

To investigate the phase-space structure of the system, it is convenient to write the normal form as
\be\label{HDZM}
K^{(1:1)} = J_1 +  J_2 +  \varepsilon^2
\left[\delta J_{1} - \frac{3}{8} \left(q J_1^2 + \frac{1}{q} J_2^2 + \frac23 J_1 J_2 ( 2 + \cos(2 \theta_1 - 2 \theta_2) ) \right)\right]\ee
where the action--angle variables are introduced according to
\ba\label{AAV}
X &=& \sqrt{2 J_1} \cos \theta_1,\\
P_X &=& \sqrt{2 J_1} \sin \theta_1,\\
Y &=& \sqrt{2 J_2} \cos \theta_2,\\
P_Y &=& \sqrt{2 J_2} \sin \theta_2.\ea
The structure of (\ref{HDZM}) displays the effect of the symmetries on the resonant part: angles appear only through the combination $2 \theta_1 - 2 \theta_2$ and this shows why the symmetric 1:1 resonance can also be dubbed a ``2:2'' resonance. We now have an integrable system with dynamics generated by Hamiltonian (\ref{HDZM}) with a second independent integral of motion 
\be\label{SIM}
 {\cal H}_0 \equiv {\cal E}=J_1+J_2.\ee
 This system is the simplest integrable approximation of the non-integrable dynamics in the logarithmic potential dominated by the lowest order resonance. In the normal form  (\ref{HDZM}), the perturbation parameter appears as a remind to denote terms of the same order and, according to the criterion stated above (see subsection 2.1), is put equal to unity in the computations. To get a quick overview of its structure we can use (\ref{HDZM}) to identify the main periodic orbits. The procedure is the following (Sanders and Verhulst, 1985, sect.7.4): we perform the canonical transformation to ``adapted resonance coordinates''
\ba
\psi &=& 2 (\theta_1 - \theta_2), \label{psi}\\
\chi &=& 2 (\theta_1 + \theta_2), \label{chi}\\
J_1 &=& ({\cal E} + {\cal R})/2, \label{j1}\\
J_2 &=& ({\cal E} - {\cal R})/2, \label{j2}\ea
where $\cal E$ is defined in (\ref{SIM}) and 
\be\label{DIM}
 {\cal R}=J_1-J_2.\ee
Since $\chi$ is cyclic and its conjugate action ${\cal E}$ is the additional integral of motion, we introduce the ``effective'' Hamiltonian
\be
{\widetilde K} = K^{(1:1)}  ({\cal R}, \psi ; {\cal E},q) ,\ee 
namely
\be\label{KER}
{\widetilde K} =  {\cal E} + 
\frac12 (q-1) ({\cal E} + {\cal R}) + A({\cal E}^{2} + {\cal R}^{2}) + B{\cal E}{\cal R} +C ({\cal E}^{2} - {\cal R}^{2})  (2 + \cos \psi), \ee
with 
\ba
A &=& -\frac{3(q^{2}+1)}{32 q}  ,\label{muA}\\
B &=& -\frac{3(q^{2}-1)}{16 q} , \label{muB}\\
C &=& -\frac{1}{16}.\label{muC}\ea
Considering the dynamics at a fixed value of ${\cal E}$, we have that ${\widetilde K}$ defines a one--degree of freedom $(\psi,{\cal R})$ system with the following equations of motion
\ba
{\dot \psi} &=& {\widetilde K}_{\cal R} = 
\frac12 (q-1) + B{\cal E} +2\left(A - C (2 + \cos \psi)\right){\cal R},\label{dpsi}\\
{\dot {\cal R}} &=& - {\widetilde K}_{\psi} = 
C \left({\cal E}^{2}  - {\cal R}^{2} \right) \sin \psi.\label{dr}
\ea
Let us determine the fixed points of this system: these in turn give the periodic orbits of the original system. The right hand side of (\ref{dr}) vanishes in one of the following three cases: 

\noindent
I. For ${\cal R}=\pm {\cal E}$;

\noindent
II. For $\psi = 0$;

\noindent
III. For $\psi = \pm \pi$. 

In the first case, the right hand side of (\ref{dpsi}) vanishes when
\be\label{psiI}
1 - q - 2 \left[B \pm 2 \left(A - C (2 + \cos \psi)\right) \right]{\cal E}=0\ee
and the two periodic orbits
\ba
&& {\cal R}={\cal E}, \;\;\;\;  J_{2}=0, \quad ({\rm Type \; Ia}), \label{Ia}\\ 
&& {\cal R}=-{\cal E}, \; J_{1}=0, \quad ({\rm Type \; Ib}), \label{Ib}\ea
ensue. The orbit of Type Ia is the periodic orbit along the $x$-axis (long axial orbit) whereas the orbit of Type Ib is the periodic orbit along the $y$-axis (short axial orbit).

In the second case, the right hand of (\ref{dpsi}) vanishes when
\be\label{psizero}
{\cal R} = \frac{q-1 + 2B  {\cal E} }{4 (3C-A)} = 
                \frac{8 q - 3 (1+q) {\cal E}}{3(q-1)}, \;\;(\psi=0)\ee
where (\ref{muA}--\ref{muC}) have been used. This fixed point determines the ``inclined'' orbit
\be
J_1 = \frac{4 q - 3 {\cal E}}{3(q-1)}, \; 
J_2 = \frac{q (3 {\cal E}- 4)}{3(q-1)}, \; ({\rm Type \; II}).\ee
Note that 
\be\label{range}
0 \le J_1,J_{2} \le {\cal E}\ee
and this range determines the condition for existence of the orbit of Type II. 

In the third case, the right hand of (\ref{dpsi}) vanishes when
\be\label{psipi}
{\cal R} = \frac{q-1 + 2B {\cal E} }{4 (C-A)} = 
               \frac{8 q (q-1) + 3 (1-q^{2}) {\cal E}}{3 q^{2}-2q+3}, \;\;(\psi=\pi).\ee
The fixed point in (\ref{psipi}) determines the elliptic (``loop'') orbit
\be\label{LOOP}
J_1 = \frac{(3-q) {\cal E} + 4 q (q-1)}{3 q^{2}-2q+3}, \; 
J_2 = \frac{q(3q-1) {\cal E} - 4 q (q-1)}{3 q^{2}-2q+3}, \; ({\rm Type \; III}).\ee
The range (\ref{range}) still determines the condition for existence of the orbit of Type III. 

\subsection{Stability analysis of the 1:1 resonance}

Let us now consider the question of the stability of the periodic orbits. As discussed in the previous section, the 1:1 resonant normal form essentially captures those features of the system characterized by the lowest order of commensurability between frequencies. In particular, it is able to describe the cases in which the nominal frequency characterizing a periodic orbit happens to become equal to that of a normal perturbation. The characteristic curve representing this equivalence in some suitable parameter space provides the stability--instability transition looked for.

For orbits of Types II and III, an ordinary investigation of the equations of variations of the system (\ref{dpsi},\ref{dr}) is enough to perform the linear stability analysis (Contopoulos, 1978) in analogy with the Floquet method. However, in the case of axial orbits of Type I, action--angle variables have singularities on them and these affect also the adapted resonance coordinates. However, the remedy is quite straightforward: to use a mixed combination of action--angle variables on the orbit itself and Cartesian variables for the other degree of freedom. 

Let us start with the orbits of Type II and III that, due to their non-singularity, are usually referred to as {\it periodic orbits in general position}. The system of differential equations for the perturbations $(\delta \psi, \delta {\cal R})$ is given by
\be
\frac{d}{dt} \left( \begin{array}{c} \delta \psi \\ \delta {\cal R} \end{array} \right) = 
\left( \begin{array}{cc} {\widetilde K}_{{\cal R} \psi} & {\widetilde K}_{{\cal R}{\cal R}} \\ 
-{\widetilde K}_{\psi \psi} &-{\widetilde K}_{{\cal R} \psi} \end{array} \right)
\left( \begin{array}{c} \delta \psi \\ \delta {\cal R} \end{array} \right).\ee
The sign of the determinant of the Hessian matrix computed on the periodic orbit, $PO$,
\be
\Delta =  ({\widetilde K}_{{\cal R} \psi}^{2}-{\widetilde K}_{{\cal R} {\cal R}}{\widetilde K}_{\psi \psi}) \vert_{PO}\ee
determines the fate of a perturbation: if $\Delta$ is negative it gives the {\it frequency} of bounded oscillating solutions
\be\label{pertfreq}
\omega_{p}=\frac1{q}\sqrt{-\Delta},\ee
where the factor $1/q$ is due to the rescaling (\ref{newE}). 
If $\Delta$ is positive it gives the characteristic exponent of the time evolution of a growing perturbation  
\be\label{fold}
\tau = \pm \frac1{q} \sqrt{\Delta}.\ee
Therefore, with our 1:1 resonant normal form (\ref{KER}), the condition for stability is 
\be\label{DETE}
\Delta =  4 C^2 {\cal R}^2 \sin^2 \psi - 2 
        C ({\cal E}^2 + {\cal R}^2) \cos \psi(A - C (2 + \cos \psi)) \vert_{PO} < 0.\ee
          Observing that each periodic orbit is identified by a fixed value of the pair $(\psi, {\cal R})$ and using the values of the constants in (\ref{muA}) and (\ref{muC}), we see that the parameter space is spanned by the ellipticity $q$ and the second integral of motion ${\cal E}$: usually, characteristic curves in this space provide the instability threshold. We remark that the choice of the integral ${\cal E}$ as one of the coordinates in the parameter space is the simplest one and is usually adopted for a qualitative analysis (Sanders and Verhulst, 1985): however, for quantitative predictions and especially for comparisons with other approaches, a more natural choice would be the physical value of the energy $E$, of which ${\cal E}$ (but for a rescaling) is only a first order approximation. In general, using $E$ in this framework is usually quite difficult for computational problems: this is one of the problems we mentioned above with exploiting the normal form in the normalizing coordinates. Using $E$ with the original Hamiltonian and an approximate integral of motion is instead not only natural but also computationally easier. At the end of the subsection we will see how to overcome the problem in the particular case of orbits of Type I and how to express the characteristic curves in terms of $E$.
          
An immediate application of condition (\ref{DETE}) concerns the inclined and loop orbits discussed above. In the cases II and III, we get respectively
\be\label{DETERII}
\Delta \vert_{\psi = 0} = 2 \left({\cal E}^{2}  - {\cal R}^{2} \right) C (A-3C) \;\; ({\rm Type \; II}) \ee
and
\be\label{DETERIII}
\Delta \vert_{\psi = \pi} = 2 \left({\cal E}^{2}  - {\cal R}^{2} \right) C (C-A) \;\; ({\rm Type \; III}).\ee
From the obvious inequality ${\cal E} > {\cal R}$,  $\Delta \vert_{\psi = 0}$ is always positive and $\Delta \vert_{\psi = \pi}$ is always negative: therefore the inclined (type II) orbit is unstable for every values of $q$ and ${\cal E}$, whereas the loop (type III) orbit (when it exists, compare with conditions (\ref{LOOP}) and (\ref{range})) is stable for every values of $q$ and ${\cal E}$.

Our major concern is however to analyze the stability properties of Type I orbits, the axial orbits, also denoted as {\it normal modes} because each of them is identified by only one of the actions. Using action--angle variables on the normal mode and Cartesian variables on the normal bundle to it, the ensuing procedure is then first to determine the condition for the normal mode to be a critical curve of the Hamiltonian in these coordinates. Second, to assess its nature (Kummer, 1977; Contopoulos, 1978; Sanders and Verhulst, 1985, sect.7.4.4): the condition is found by considering the function
\be\label{LM1}
K^{(\mu)}=K+\mu {\cal H}_0,\ee
where $\mu$ has to be considered as a {\it Lagrange multiplier} to take into account that there is the constraint $ {\cal H}_0 = {\cal E}$. The Lagrange multiplier is found by imposing
\be\label{LM2}
{\rm d} \, K^{(\mu)} =0,\ee
that is the total differential of (\ref{LM1}) vanishes on the normal mode. Its nature is assessed by computing the matrix of second derivatives of $K^{(\mu)}$: if the Hessian determinant of the second variation is positive definite the mode is elliptic stable; if it is negative definite the mode is hyperbolic unstable. In the case of the {\it y}-axis orbit of Eq.(\ref{Ib}), good coordinates are $X,P_X$ and
\ba\label{AAW1}
J &=& J_{2},\\
\theta &=& \theta_{2},\ea
so that the periodic orbit is given by 
\be\label{TIb}
X=P_X=0, \;\; J={\cal E}.\ee 
The terms in the normal form are then
\be
{\cal H}_{0} = \frac12 (X^{2}+P_X^{2})+J\ee
and
\be\begin{array}{ll}
K_{2} = & \frac12 (q-1) (X^{2}+ P_X ^{2}) - \frac{3}{32} \left[q(X^{2}+ P_X^{2})^2 + \frac4{q} J^2 \right] -\\  
      & \frac18 J \left[ 2 (X^{2}+ P_X^{2}) + (X^{2}-P_X^{2}) \cos 2\theta + 2 X P_X \sin 2\theta \right].        \end{array}\ee
It is straightforward to check that, in this case, Eq.(\ref{LM2}) reduced to the periodic orbit defined by  
Eq.(\ref{TIb}) gives
\be
\mu + 1 - \frac{3 {\cal E}}{4q} = 0,\ee
which allows us to find the required value of the Lagrange multiplier. With this result, the matrix of the second derivatives of $K^{(\mu)}$ on the periodic orbit is
\be \frac18
\left( \begin{array}{cc} 4(q-1) - {\cal E} [ (2+\cos 2\theta) - (3/q)] & 
                                   - {\cal E} \sin 2\theta \\ 
                                   - {\cal E} \sin 2\theta &
                                   4(q-1) - {\cal E} [(2-\cos 2\theta) - (3/q)] \end{array} \right).\ee
The equation ${\rm det}[d^2 K^{(\mu)}({\cal E})]=0$ gives
\be\label{DETER11}
(q-1)(4 q- 3 {\cal E})(4 q^2 + 3 {\cal E} - q (4+{\cal E}))=0\ee
and, according to the above recipe, for stability this polynomial in $(q,{\cal E})$ must be positive. With $q$ in the range (\ref{rangeq}) as the independent parameter, the instability condition is 
\be\label{diseDZMY}
\frac{4 q (1-q) }{3-q}
<{\cal E}<
\frac{4q}{3}.
\ee
It is interesting to remark that the instability threshold corresponding to the first inequality in (\ref{diseDZMY}) coincides with the condition of existence of the loop orbit. In fact, from the obvious condition (\ref{range}) and the values in (\ref{LOOP}), we see that their validity is satisfied exactly by the first inequality in (\ref{diseDZMY}): this is equivalent to state that at the point in the parameter space where the short axis orbits becomes unstable, the sequence of loop orbits bifurcates. The second inequality corresponds to a return to stability that actually disappears with the higher order treatment.

Proceeding in the same way in the case of the {\it x}-axis orbit of Eq.(\ref{Ia}), analogous expressions can be obtained using non-singular coordinates $J, \theta, Y, P_{Y}$ and it is quite straightforward to check that, at this level of approximation, the Type Ia orbit is always stable. The procedure above shows the amount of information that can be extracted from the 1:1 resonant normal form. In this respect, it is analogous to the first order averaging approach applied by de Zeeuw and Merritt (1983) in studying the Schwarzschild potential. 

Before going forward, we want to discuss how to give a more concrete meaning to the result enbodied by (\ref{diseDZMY}): this is possible if the physical energy $E$ appears explicitly. According to the rescaling (\ref{newE}), we assume that 
\be
\frac{n E}{\omega_2} \ee
is the constant `energy' value assumed by the truncated Hamiltonian $K$. In the present instance, $n=1$ and $ \omega_2=1/q$, so that, on the {\it y}-axis orbit given by (\ref{TIb}), we have
\be\label{KIb}
K = {\cal E} - \frac{3}{8 q} {\cal E}^{2} + ... = qE.\ee
The dots are present to recall that a remainder has been neglected. The series (\ref{KIb}) can be inverted to give
\be
{\cal E}= q (E + \frac38 E^{2}+ ...)\ee
and this can be used in the treatment of stability to replace ${\cal E}$ with $E$. With this substitution, the critical curve corresponding to the first inequality in (\ref{diseDZMY}) is simply
\be\label{Eq11}
E (q) = 2(1-q).\ee
Expression (\ref{KIb}) is also useful to find the oscillation frequency on the periodic orbit,
\be\label{freq11}
\kappa_{2}= \frac1{q} \frac{\partial K}{\partial {\cal E}} = 
                    \frac1{q} \left( 1 - \frac{3}{4 q} {\cal E} \right),\ee
which is needed if one is interested in computing either the {\it frequency ratio}
\be\label{freqrat11}
r_2=\frac{\omega_{p}}{\kappa_{2}},\ee
where the perturbation frequency $\omega_{p}$ is defined in (\ref{pertfreq}), or the {\it e-folding rate}
\be\label{efold11}
\tau_2=\frac{\tau}{\kappa_{2}},\ee
where the time scale $\tau$ is defined in (\ref{fold}).

The analysis with a normal form truncated at higher order would provide more accurate predictions. However, for the time being, we want first to make the analogous analysis with the approximate integral, to better understand the relationship between the two approaches.

\subsection{Stability analysis with the approximate integral in the original variables}

Recalling the generic expression of the terms in the integral of motion (\ref{eqn:In}), if we truncate to second order, we have
\be\label{inva2}
I^{(1:1)} = H_{0}+\varepsilon^2(V_{2}-K_{2}^{(1:1)}).\ee
This is the best approximate integral of motion of a symmetric perturbed 1:1 oscillator to order $\varepsilon^4$ in the perturbation parameter: in fact, due to the symmetries of the problem, odd-degree terms are absent both in the normal form and in the approximate integral. A remark on how to interpret terms in (\ref{inva2}) is necessary: even if not explicitly indicated, all of them must be intended as polynomials in the original variables $(p_x, p_y, x, y)$. In particular, $K_{2}^{(1:1)}$ is the same as  in (\ref{K211}), where capital letter variables are simply replaced by the corresponding lower case letters.

To assess the stability of periodic orbits, we may proceed in the following way: starting for definiteness with the {\it y}-axis orbit, to use $I^{(1:1)}$ and the conserved energy $E$ to construct an $x-p_x$ Poincar\'e section by means of the intersection of the function $I^{(1:1)} (x,y,p_x;E)$ with the $y=0$ hyperplane. The level curves of the function
\be\label{FF}
F = I^{(1:1)} (x,0,p_x;E)\ee
allow us to determine the nature of critical points and invariant curves. Critical points ($CP$) on the surface correspond to periodic orbits in phase space (Contopoulos, 2002) and their nature gives the necessary information: they are either extrema (elliptic fixed point = stable periodic orbit) or saddles (hyperbolic  fixed point = unstable periodic orbit) and this can be assessed by using the Hessian determinant
\be (F_{p_x p_x} F_{x x} - F_{x p_x}^{2}) \vert_{CP}.\ee 
Clearly, for a periodic orbit, even the location of the critical points can be already quite difficult and this limits the generality of the approach. However, in the case of axial orbits, the approach is straightforward: the {\it y}-axis orbit, for example, coincides with the origin in this section. The second derivatives at the origin are 
\ba
F_{p_x p_x} \vert_{CP} &=& 4 (q-1),\\
F_{x x} \vert_{CP} &=& 4 (q-1) + 2 E,\\
F_{x p_x} \vert_{CP} &=& 0 .\ea
The range of instability (namely, the range where the two non vanishing second derivatives have different sign) is
\be\label{diseINV}
E > 2 (1-q),
\ee
in agreement with (\ref{Eq11}). We remark that the results reported in Belmonte et al. (2006, cfr. Table 1), were obtained just by using both (\ref{diseDZMY}) and (\ref{diseINV}) pointing out the discrepancy between the two predictions. An analogous procedure can be followed for the {\it x}-axis orbit by constructing a $y-p_y$ Poincar\'e section and studying the level curves of the function
\be\label{FFY}
F = I^{(1:1)} (0,y,p_y;E)\ee 
obtained by means of the intersection of the function $I^{(1:1)} (x,y,p_y;E)$ with the $x=0$ hyperplane. In agreement with the result of the previous subsection, the {\it x}-axis orbit is predicted to be stable using the present critical point analysis.


\section{1:2 symmetric resonance and fourth order normalization}

The loop orbit bifurcates from the {\it y}-axis orbit at its instability threshold and is correctly described by the 1:1 (detuned) resonant normal form because its appearance coincides with the equality of the frequencies of the axial orbit and that of a normal perturbation. The `banana' orbits  bifurcate from the {\it x}-axis when the frequency of the axial orbit falls to {\it one half} of that of a normal perturbation. This is the reason why the 1:1 resonant normal form does not detect any change of stability and we need the 1:2 (detuned) resonant normal form. A detailed study of the orbits in this resonance has been provided by Contopoulos (1963).

\subsection{Orbit structure of the 1:2 resonance}

In the case of the $m=1,n=2$ resonance in presence of reflection symmetries about both axes, we know that the normal form must be pushed at least to order 4, since it has to include terms of degree $2 \times (m+n) = 6$. Therefore, we have to perform a further step of normalization to include $K_4$ in the normal form and the system to solve is now
\ba\label{EHK4}
&& K_2=V_2+M_{2}{\cal H}_{0},\\
&& K_4=V_4+M_{2}V_2+M_{4}{\cal H}_{0}.\ea
The expression of the normal form is quite involved (cfr. Belmonte et al. 2006), but we can exploit the change of variables to action--angle coordinates as defined in (\ref{AAV}) to see more easily its structure:
 \be\label{HDZM12}
K^{(1:2)}  = J_1 +  2 J_2 + \varepsilon^2 \bigl(2 \delta J_1 - P_2(J_1 , J_2)\bigr) + \varepsilon^4 \bigl(P_3 (J_1 , J_2) + \frac98 J_1^2 J_2 \cos(4 \theta_1 - 2 \theta_2)\bigr),\ee
where the polynomials $P_2$ and $P_3$ are homogeneous of degree $2$ and $3$ respectively,
\begin{eqnarray*}
P_{2} &=& \frac34 \left(q J_{1}^2 + \frac1{q} J_{2}^2 \right) + J_{1} J_{2},    \\
P_{3} &=& q \left(\frac56 - \frac{17}{16} q \right) J_{1}^3 + \left(\frac{13}{12} - \frac32 q \right) J_{1}^2 J_{2} - \left(\frac5{12} - \frac{3}{4q} \right) J_{1} J_{2}^2 + \frac{29}{96} J_{2}^3. 
\end{eqnarray*}
and the frequency ratio (\ref{DET}) now is
\be\label{DET12}
\frac{\omega_1}{\omega_2} = \frac12 + \varepsilon^2 \delta.\ee
The adapted resonance coordinates are now defined by
\ba
\psi &=& 4 \theta_1 - 2 \theta_2, \label{psi2}\\
\chi &=& 4 \theta_1 + 2 \theta_2, \label{chi2}\\
J_1 &=& {\cal E} + 2 {\cal R}, \label{j12}\\
J_2 &=& 2 {\cal E} - {\cal R}. \label{j22}\ea
As before, knowledge of the orbit structure is obtained by investigating the fixed points of the one-dimensional dynamical system associated to the effective Hamiltonian
\be
{\widetilde K}^{(1:2)} = K^{(1:2)}  ({\cal R}, \psi ; {\cal E},q) ,\ee
where
\be
{\cal R}=\frac15 (2J_1-J_2),\ee
at fixed values of the integral of motion
\be\label{E12}
{\cal E}=\frac15 (J_1+2J_2).\ee

The study of existence and stability of periodic orbits proceeds in the same way as in the previous section. In the present case the normal modes (axial orbits) are given by
\ba
 {\cal R}&=&{2\cal E}, \;\;\;\;  J_{2}=0, \quad ({\rm Type \; Ia}), \label{IIa}\\ 
 {\cal R}&=&-{\cal E}/2, \; J_{1}=0, \quad ({\rm Type \; Ib}). \label{IIb}\ea
The other periodic orbits are identified by the zeros of the system
\ba
&& {\widetilde K}^{(1:2)}_{\cal R} =0, \label{dR2}\\ 
&& {\widetilde K}^{(1:2)}_{\psi} =0. \label{dpsi2}\ea
In one case, (\ref{dpsi2}) is satisfied by $\psi=0$ and the corresponding root of (\ref{dR2})
\be\label{psizero2}
{\cal R} = {\cal R}_{B}({\cal E},q), J_1 = J_{1(B)} , J_2 = J_{2(B)},\quad \ee
determines the ``banana'' orbit. In the other case, (\ref{dpsi2}) is satisfied by $\psi=\pi$ and the corresponding root of (\ref{dR2})
\be\label{psizeropi}
{\cal R} = {\cal R}_{A}({\cal E},q), J_1 = J_{1(A)} , J_2 = J_{2(A)},\ee
determines the ``antibanana''. From (\ref{E12}), their range is given by
\be\label{range12B}
0 \le J_{1(B,A)} \equiv 2 {\cal R}_{B,A}({\cal E},q) + {\cal E} \le 5{\cal E}\ee
and
\be\label{range12A}
 0 \le  J_{2(B,A)} \equiv 2 {\cal E} - {\cal R}_{B,A}({\cal E},q) \le \frac52 {\cal E}.\ee

\subsection{Stability analysis with the normal form}

The procedure to determine the condition of stability of axial orbits leads to analyze critical curves of the modified function (\ref{LM1}) where $K$ is now given by (\ref{HDZM12}). From the analysis of the 1:1 resonance, we have seen that it cannot detect any instability threshold of the long axis periodic orbit and therefore we readily address this question in the framework of the 1:2 resonance. Considering the {\it x}-axis (type Ia, Eq.(\ref{IIa})) orbit, good coordinates are given by
\ba\label{AAW2}
X &=& \sqrt{2 J} \cos \theta,\\
P_X &=& \sqrt{2 J} \sin  \theta,\\
Y &=& Y,\\
P_Y &=& V,\ea
so that the periodic orbit is given by 
\be\label{TIIb}
Y=V=0, \;\; J=5{\cal E},\ee 
 and
\be\label{LM12}
{\cal H}_{0} = J + Y^{2} + V^{2}.\ee
Condition (\ref{LM2}) allows us to find the Lagrange multiplier
\be
\mu = \frac{1}{16}(-32 q + 24{\cal E} q - 40{\cal E}^2 q + 51 {\cal E}^2 q^2).\ee
The equation ${\rm det}[K^{(\mu)}({\cal E})]=0$, obtained by computing the matrix of the second derivatives of $K^{(\mu)}$ on the normal bundle to the periodic orbit (\ref{TIIb}), is 
\be
\label{d12X}
(48 - 96 q + 24 (3 q -1) {\cal E} + (26 - 153 q + 153  q^2) {\cal E}^2)
(48 - 96 q + 24 (3 q -1) {\cal E} + (26 - 159 q + 153  q^2) {\cal E}^2)= 0.\ee
 whose zeros are
 \ba
{\cal E}_{B\pm}&=&4{{3 - 9 q \pm \sqrt{3} \sqrt{-23 + 187 q - 432 q^2 + 306 q^3}} \over 
    {26 - 153 q + 153 q^2}},\\
{\cal E}_{A\pm}&=&4{{3 - 9 q \pm \sqrt{3} \sqrt{-23 + 193 q - 444 q^2 + 306 q^3}} \over 
    {26 - 159 q + 153 q^2}}.\ea
 It happens that the conditions of existence in (\ref{range12B},\ref{range12A}) are satisfied exactly when 
 \be\label{banrange}
     {\cal E}_{B+} \le {\cal E} \le {\cal E}_{B-}\ee
     for the banana orbit and
 \be\label{abanrange}
     {\cal E}_{A+} \le {\cal E} \le {\cal E}_{A-}\ee
     for the antibanana. We again see the relation between transition to instability of a normal mode and bifurcation of a resonant periodic orbit, since, outside the above ranges, the determinant $\Delta = {\rm det}[K^{(\mu)}({\cal E})]$ is positive. As before, we want to make the analogous analysis with the approximate integral, to better grasp the relationship between the two approaches.
      
\subsection{Stability analysis with the approximate integral of motion}

      We can develop the parallel approach of determining the nature of fixed points on the surface of section also for higher order approximate integrals. Using  (\ref{eqn:In}), if we truncate at order four, we have
\be\label{inva4}
I^{(1:2)} = H_{0}+\varepsilon^2(V_{2}-K_{2})+\varepsilon^4(V_{4}-\{g_{2},K_{2}\}-K_{4})\ee
where 
\be
g_{2} = \frac{1}{24}P_{X}^2 P_{Y} Y -\frac{3q}{16} P_{X}^3 X - \frac13 P_{X} P_{Y}^2 X- \frac{5q}{16} 
    P_{X} X^3 - \frac{3}{32 q} P_{Y}^3 Y - \frac{7}{24} P_{Y} X^2 Y - \frac16 P_{X} X Y^2 - 
    \frac{5}{32 q} P_{Y} Y^3
\ee 
is the second order generating function determined in the first step of the normalization. This is the best approximate integral of motion of a symmetric perturbed 1:2 oscillator to order $\varepsilon^6$ in the perturbation parameter. Let us again consider the {\it x}-axis orbit: the construction of the $y-p_y$ Poincar\'e section and the study of the critical points of the function
\be\label{FFFY}
F = I_4 (0,y,p_y;E),\ee 
obtained by means of the intersection of the function $I_4 (x,y,p_y;E)$ with the $x=0$ hyperplane,  proceeds as above. Concerning the nature of the fixed point in the origin, we get quadratic inequalities in the energy analogous to those arising from (\ref{d12X}): the discrepancy between the two methods  is still of order $\delta^{2}$. The locus of points where the origin in the $y-p_y$ Poincar\'e section changes from an extremum to a saddle are
   \ba
E_{B\pm}&=&4{{-8 + 10 q^2 \pm \sqrt{228 q - 887 q^2 + 830 q^3 + 100 q^4}} \over 
    {-76  + 165 q^2}},\label{EBpm}\\
E_{A\pm}&=&4{{-4 + 2 q^2 \pm \sqrt{180 q - 659 q^2 + 614 q^3 + 4 q^4}} \over 
    {15 (- 4 q + 7 q^2)}}.\label{EApm}\ea
  These are clearly related to those found above but are not the same: to reconcile the two predictions, it is first necessary to give the relation between the new `energy' $ {\cal E} $ and the physical energy $E$. But this is actually not enough: as seen in the 1:1 case, one must also perform a series expansion of the relations in the $(q,E)$ plane and truncate it at the same order of the normal form.
  
According to the rescaling (\ref{newE}), with $n=2$ and $ \omega_2=1/q$, on the {\it x}-axis orbit given by (\ref{TIIb}), we have
\be\label{KIa}
K =2q J - \frac34 q J^2 + \frac12 q \left(\frac53 - \frac{17}{4} q(q-1) \right) J^3 + ... = 2qE.\ee
The dots are present to recall that a remainder has been neglected. The series (\ref{KIa}) can be inverted to give
\be
J = 5 {\cal E}= E + \frac38 E^2 - \frac1{16} \left(\frac{13}{6} + 17 q (1-q) \right) E^3 + ...\ee
and this can be used in the treatment of stability to replace ${\cal E}$ with $E$. We limit ourselves to the lower bounds in the ranges in (\ref {banrange}) and (\ref {abanrange}), getting {\it sufficient} conditions for the bifurcation of, respectively, bananas and antibananas: 
\ba
&& E_{B+} = 8 \left(q - \scriptstyle\frac12 \right)  - \frac{20} 3 \left(q - \scriptstyle\frac12 \right)^2; \label{EB122} \\
&& E_{A+} = 8 \left(q - \scriptstyle\frac12 \right) + \frac{28}3 \left(q - \scriptstyle\frac12 \right)^2. \label{EA122} \ea
It can be checked by a long but easy computation that the above series coincide with those obtained by respectively expanding $E_{B+}$ in (\ref{EBpm}) and $E_{A+}$ in (\ref{EApm}) in $q$ around $q=1/2$, thus confirming the equivalence between the two methods. In the following section we present a more precise result obtained from a normal form truncated at higher order.

\section{Orbit structure of the logarithmic potential}

The theory discussed in the previous sections is applied to investigate the orbit structure of the logarithmic potential. In particular we are interested in simple recipes to predict the stability of the main periodic orbits. We can compare these analytical predictions with other results, mainly numerical, from the literature. In particular, we have chosen the work by Miralda-Escud\'e \& Schwarzschild (1989, MES in what follows) who made an accurate numerical exploration of the phase-space structure of the logarithmic potential: they give the existence parameter ranges of the main families of periodic orbits and the bifurcation ensuing from instability thresholds in two models (corresponding to ellipticity values $q=0.7$ and $q=0.9$) determined by solving the perturbation equations with the Floquet method. The authors use the core radius $R_{c}$ to parametrize the sequence of periodic orbits because they were interested in comparing the results with the case of the {\it singular} scale--free case $R_{c}=0$: in fact they fix the energy $E=0$ in all their computations and vary $R_{c}$ and $q$. For our purposes, it is more natural to use the energy and the ellipticity as parameters. To compare our results with that in MES the conversion 
\be\label{MESconv}
E = - \log R_{c}\ee
must be used. We have numerically computed the instability thresholds of the axial orbits in the range $0.6 < q < 1$, with results in agreement with MES where available.

Another approach to the study of the stability of axial orbits in 
the logarithmic potential has been followed by Scuflaire (1995, SC in what follows) who solved the Hill--like equation of the normal perturbation to the periodic orbit by means of a Poincar\'e--Lindstedt series expansion up to order 20. SC uses $a$, the amplitude of the axial orbit, as a parameter: the conversion to energy is given by
\ba
&& E = \frac12 \log (1+a^{2}), \label{Sconvx}\\ 
&& E = \frac12 \log (1+(a/q)^{2}), \label{Sconvy}\ea
on the {\it x}-axis and {\it y}-axis orbits respectively. The approach in SC is analytical and perturbative but, being based on the theory of nonlinear differential equations with periodic coefficients (Jordan \& Smith, 1999), is quite different from ours: it is therefore a very useful term of comparison because its results, even if restricted to axial orbits only, stem from a very high order perturbation analysis.

\subsection{Stability and bifurcation of closed orbits}

To get a higher precision than that obtained in the discussion of Sections 3 and 4, we have computed the 1:1 and 1:2 resonant normal form up to order 6 (degree 8 in the variables). A general analysis of these normal forms is quite cumbersome, but, concerning the axial orbits, they allows to get quite easily improved predictions of the stability thresholds.

\begin{figure}
\plotone{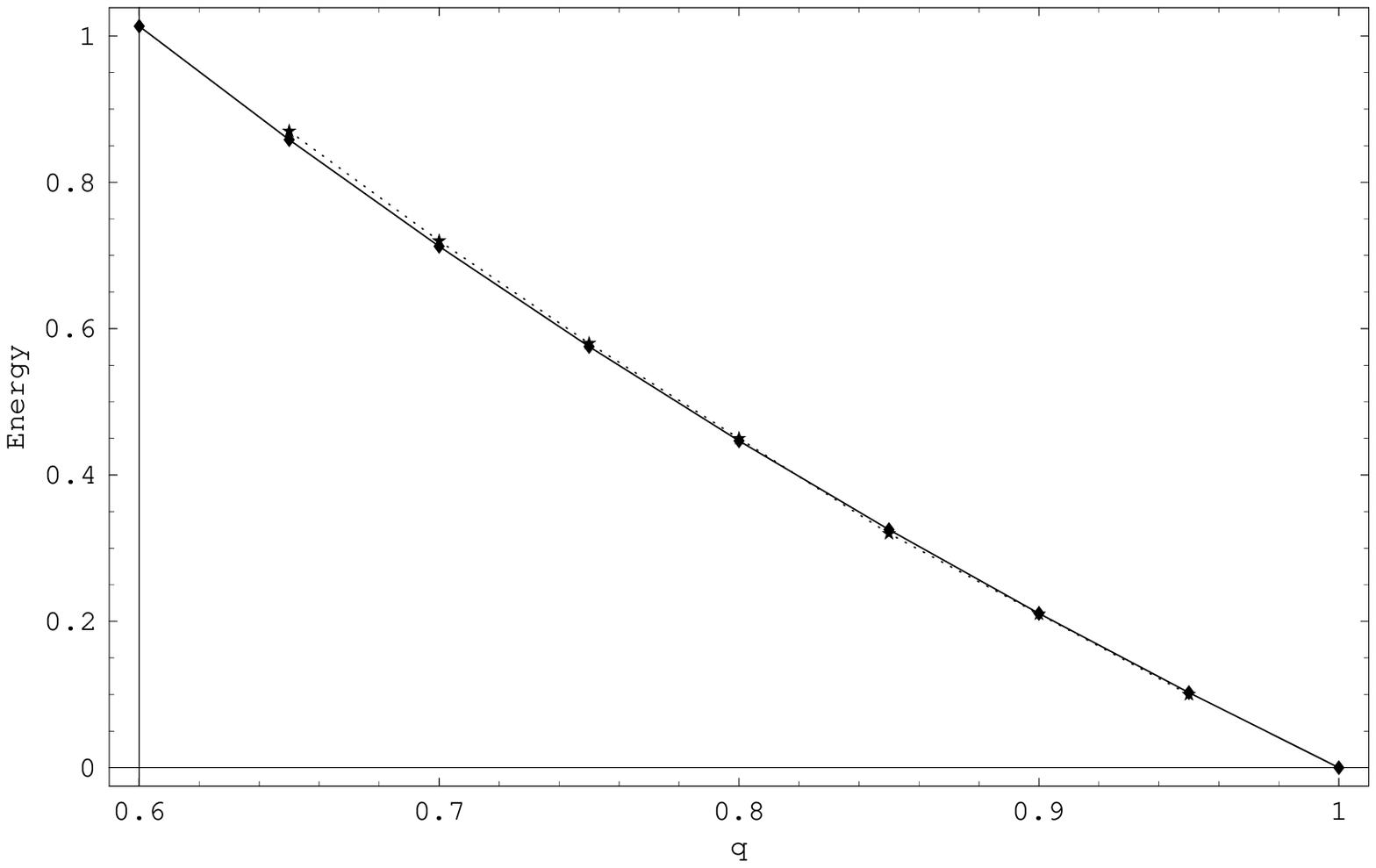}
\caption{Stability thresholds of the short  {\it y}-axis orbit: analytical (Eq. \ref{Eq611}, continuous line); numerical solution with the Floquet method (dotted line).}
\label{fig:y11}
\end{figure}

In the case of the {\it y}-axis orbit, using the same coordinate as in Subsection 3.2, the normal form on the periodic orbit $(J_1=0, J_2 =  {\cal E})$ is
\be\label{K6Ib}
K = {\cal E} - \frac{3}{8 q} {\cal E}^{2} +
                       \frac{29}{192 q^2} {\cal E}^{3} - 
                       \frac{55}{1024 q^3} {\cal E}^{4} + ... = qE.\ee
As a consequence, the upgraded evaluation of the frequency on the {\it y}-axis orbit is
\be\label{f611}
\kappa_{2} ({\cal E})=  \frac1{q} \frac{\partial K}{\partial {\cal E}} = 
                                    \frac1{q} \left(1 - \frac{3}{4 q} {\cal E} +
                       \frac{29}{64 q^2} {\cal E}^{2} - 
                       \frac{55}{256 q^3} {\cal E}^{3} \right).\ee
                       These series generalize (\ref{KIb}) and (\ref{freq11}) respectively. As before, we actually prefer to have expressions in terms of the `real' energy $E$, so we invert  (\ref{K6Ib}) to get
                       \be\label{E611}
{\cal E}= q E \left(1 + \frac38 E+  \frac{25}{192} E^{2}+ \frac{35}{1024} E^{3}+ ...\right)\ee
     and substitute in (\ref{f611}), obtaining            
       \be\label{fE611}
\kappa_{2} (E) =  \frac1{q} \left(1 - \frac{3}{4} E +
                       \frac{11}{64} E^{2} + 
                       \frac{7}{256} E^{3} \right).\ee                
Computing the determinant of the matrix of the second derivatives of the modified function (\ref{LM1}), ${\rm det}[d^2 K^{(\mu)}({\cal E})]$ and putting it equal to zero, gives
\be\label{DETER611}
(q-1)(64 q^2 - 48 q {\cal E} + (29 -17 q) {\cal E}^2)
         (64 q^2 (1+q) + 16 q (3 - q) {\cal E} -(q^2-2q+29){\cal E}^2)=0\ee
generalizing (\ref{DETER11}).
Substituting (\ref{E611}) and solving for $E$ we get
                \be\label{Eq611}
E (q) = 2(1-q)+(1-q)^2-\frac56 (1-q)^3,\ee
obtaining a prediction of the stability--instability transition of the short axis orbit (with subsequent bifurcation of the loop orbit) up to third order in the $(q,E)$ plane.                       
                       The values obtained from this relation are shown in Fig.(\ref{fig:y11}), continuous line, and are compared with the numerical results with the Floquet theory (dotted line) showing a very good agreement: the values obtained by MES (some of the dots in the figure) are $0.21$ in the case $q=0.9$ and $0.72$ in the case $q=0.7$. Moreover, the expansions (\ref{f611}) and (\ref{Eq611}) coincide up to the same order with those reported in SC when the conversion from amplitude to energy is performed using (\ref{Sconvy}). 
                       
\begin{figure}
\plotone{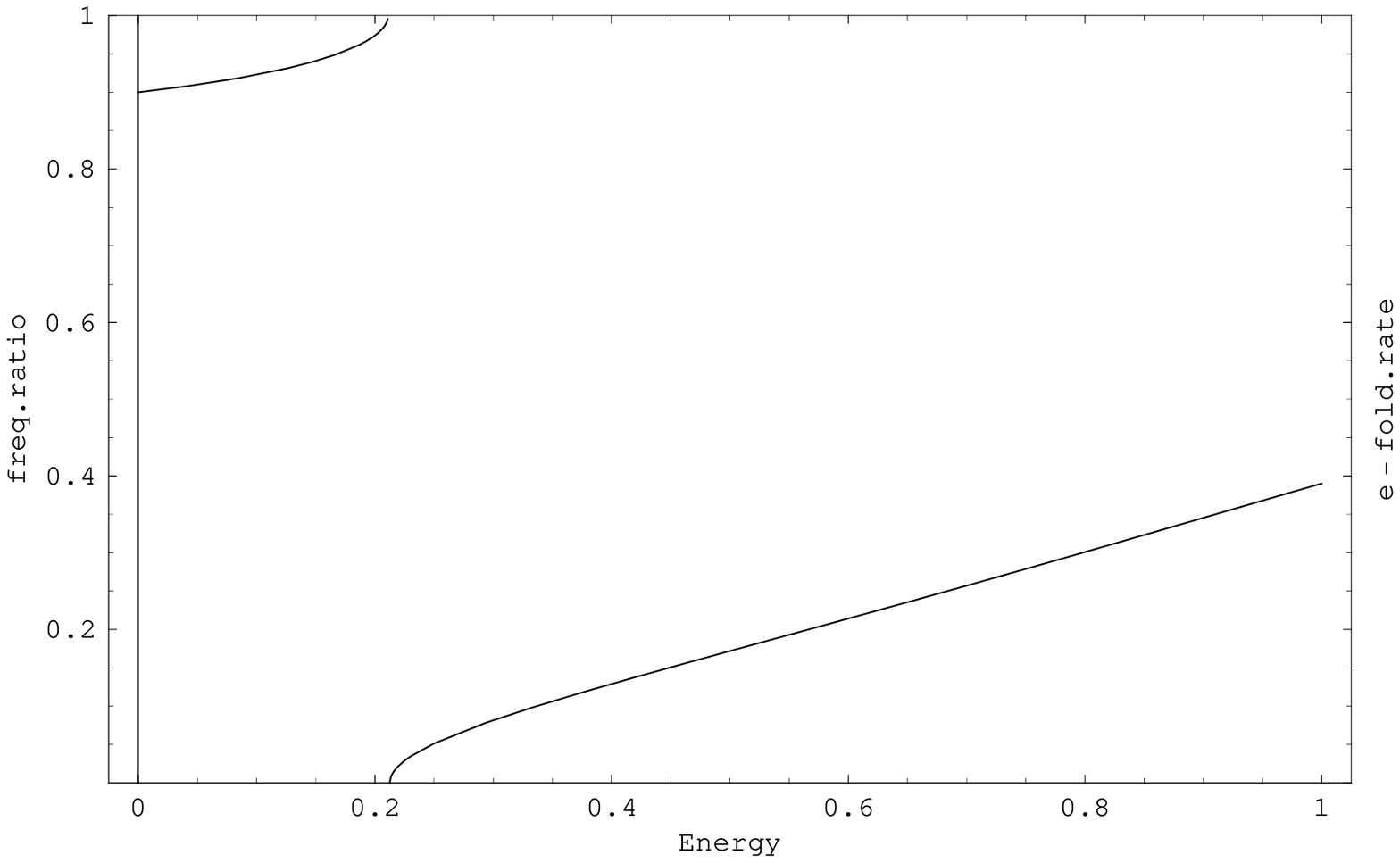}
\caption{Frequency ratio and e-folding rate for the short {\it y}-axis orbit at $q= 0.9$.}
\label{fig:110_9}
\end{figure}

\begin{figure}
\plotone{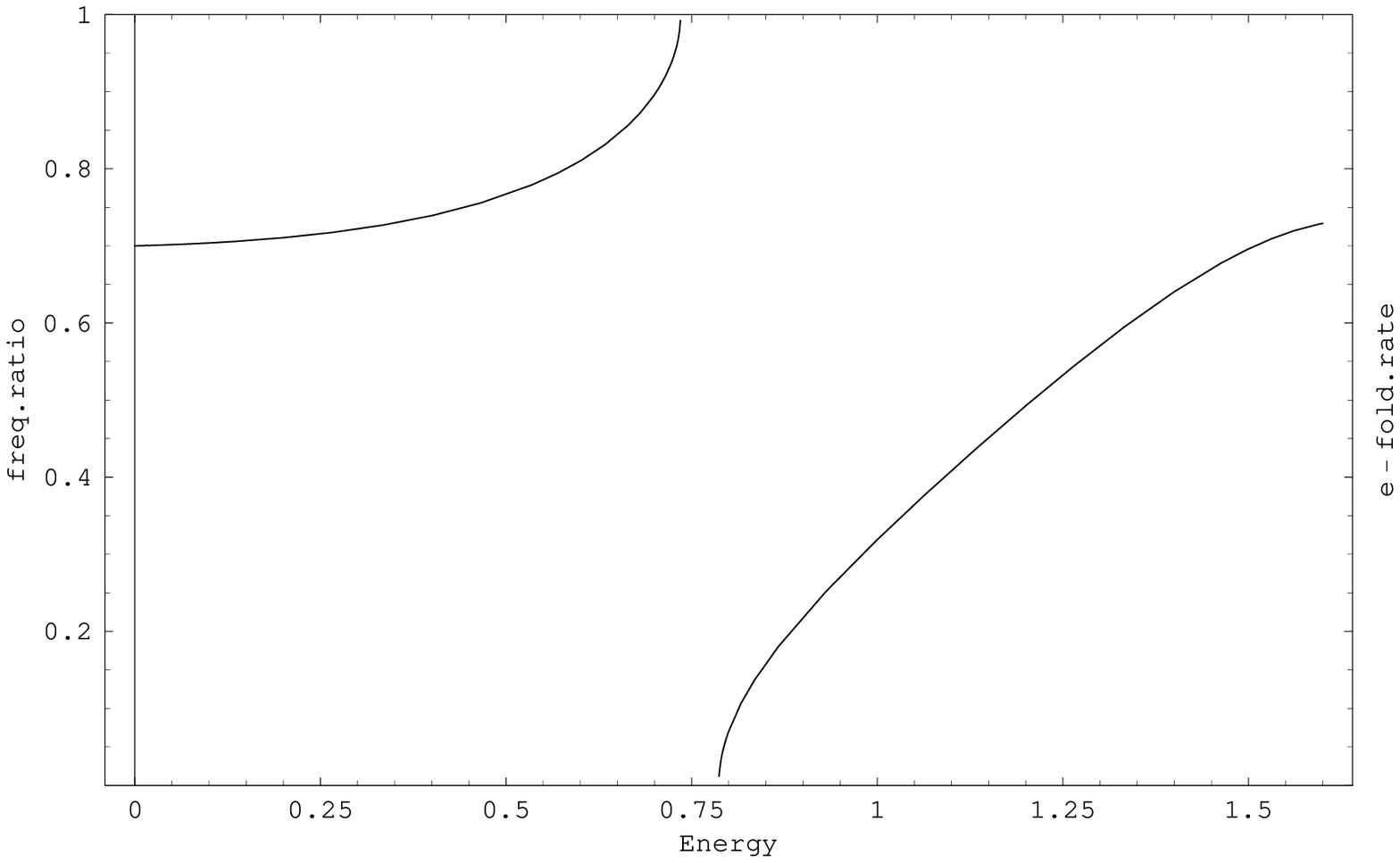}
\caption{Frequency ratio and e-folding rate for the short  {\it y}-axis orbit at $q= 0.7$.}
\label{fig:110_7}
\end{figure}

                       We may also compute the ratio (\ref{freqrat11}) between the frequency of the perturbation and that of the periodic orbit, where the perturbation frequency is associated to the determinant in the left side of (\ref{DETER611}) like in (\ref{pertfreq}) and the e-folding rate as defined in (\ref{efold11}): as expected, at low energy, the ratio has limit $q$, since $\omega_{p} \rightarrow 1, \kappa_{2} \rightarrow 1/q$, whereas, at the transition, the ratio coincides with the resonance value 1. In Fig.(\ref{fig:110_9}) the two quantities are plotted for $q=0.9$, in Fig.(\ref{fig:110_7}) they are plotted for $q=0.7$. They can be compared with the corresponding figures in MES (bottom panels in their figg.7 and 8) where they have been obtained numerically. The return to stability at higher energies mentioned in Sect. 3 (cfr. the second inequality in (\ref{diseDZMY})) does not occur in the logarithmic potential. This incompleteness appears also in the treatment made by de Zeeuw \& Merritt (1983) of the Schwarzschild potential using first order averaging. This problem is solved by our higher order approach, because the other roots of (\ref{DETER611}) are either negative or complex.
                       
\begin{figure}
\plotone{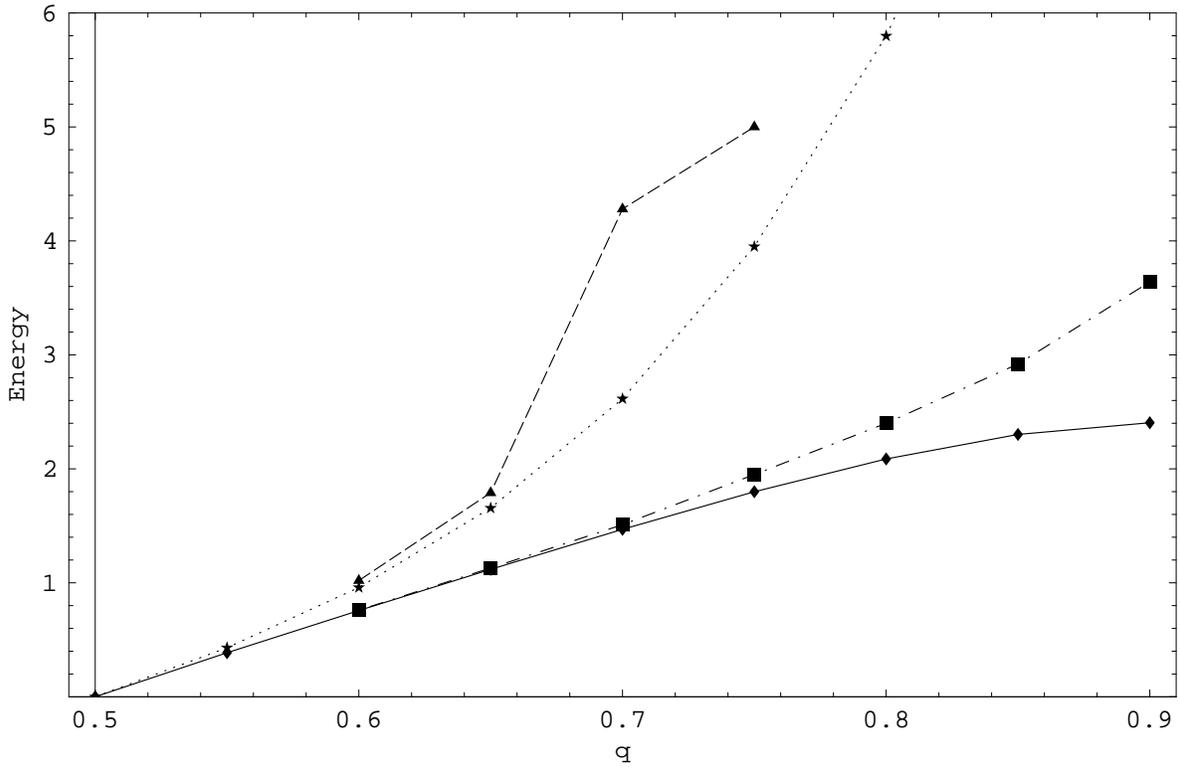}
\caption{Stability thresholds of the long  {\it x}-axis orbit: analytical (banana bifurcation, Eq. \ref{EB123}, continuous line and antibanana bifurcation, Eq. \ref{EA123}, dotted line) and numerical (dash-dotted and dashed line respectively).}
\label{fig:x12}
\end{figure}

       In the case of the {\it x}-axis orbit, we limit ourselves to provide results without derivation. The 1:2 normal form on the periodic orbit (Eq. (\ref{HDZM12}) with $J_2 = 0$) is
\be\label{K6Ia}
K =2q J_1 - \frac34 q J_1^2 + \frac12 q \left(\frac53 - \frac{17}{4} q(q-1) \right) J_1^3 -  
     \frac5{16}q \left(\frac72 - 11 q - \frac{75}{8} q^2 \right) J_1^4.\ee
As a consequence, the upgraded evaluation of the expansion of the action in terms of the true energy is
\be\label{E612}
J_{1}= E + \frac38 E^2 - \frac1{16} \left(\frac{13}{6} + 17 q (1-q) \right) E^3 + 
                    \frac5{128}\left(\frac34 + 7 q - \frac{27}2 q^{2} \right) E^4.\ee
The frequency on the {\it x}-axis orbit computed through
\be
\kappa_{1} (J_1)= \frac1{2q} \frac{\partial K}{\partial J_1} ,\ee
and expressed in term of the upgraded expansion (\ref{E612}), is
\be\label{f612}
\kappa_{1} = 1 - \frac{3}{4} E +
                          \frac{11}{64} E^{2} + 
                          \frac{7}{256} E^{3}.
\ee
Computing the determinant of the matrix of the second derivatives of the modified function (\ref{LM1}) on the periodic orbit, $\Delta = {\rm det}[d^2 K^{(\mu)}]$ and putting it equal to zero, gives, in this case, an equation of sixth degree: the two roots in addition to those in (\ref{EBpm}) and (\ref{EApm}), in analogy to the roots $E_{B-}$ and $E_{A-}$, does not put further constraints on the conditions sufficient for instability of the axial orbit. Therefore, we use the upgraded version of the relevant roots and express them as series in powers of $q - \scriptstyle\frac12$, to get 
\ba
&& E_{B+} = 8 \left(q - \scriptstyle\frac12 \right)  - \frac{20} 3 \left(q - \scriptstyle\frac12 \right)^2 + \frac{268}{9} \left(q - \scriptstyle\frac12 \right)^{3}, \label{EB123} \\
&& E_{A+} = 8 \left(q - \scriptstyle\frac12 \right) + \frac{28}3 \left(q - \scriptstyle\frac12 \right)^2 + \frac{460}{9} \left(q - \scriptstyle\frac12 \right)^{3}, \label{EA123} 
\ea
obtaining a prediction of the stability--instability transition of the long axis orbit (with subsequent bifurcation of the banana orbit) up to third order in the $(q,E)$ plane. The values obtained from these relations are shown in Fig.(\ref{fig:x12}): the continuous and dotted lines are respectively the relations in (\ref{EB123}) and (\ref{EA123}), and are compared with the numerical results with the Floquet theory, namely bifurcation of the banana (dash-dotted line) and bifurcation of antibanana (dashed line). The values obtained by MES are respectively $E=1.52$  and $E=4.29$ in the case $q=0.7$, whereas, in the case $q=0.9$, only the first transition at $E=3.62$ is available. Moreover, the expansions (\ref{f612}--\ref{EA123}) coincide up to the same order with those reported in SC when the conversion from amplitude to energy is performed using (\ref{Sconvx}). 
                       
\begin{figure}
   \plotone{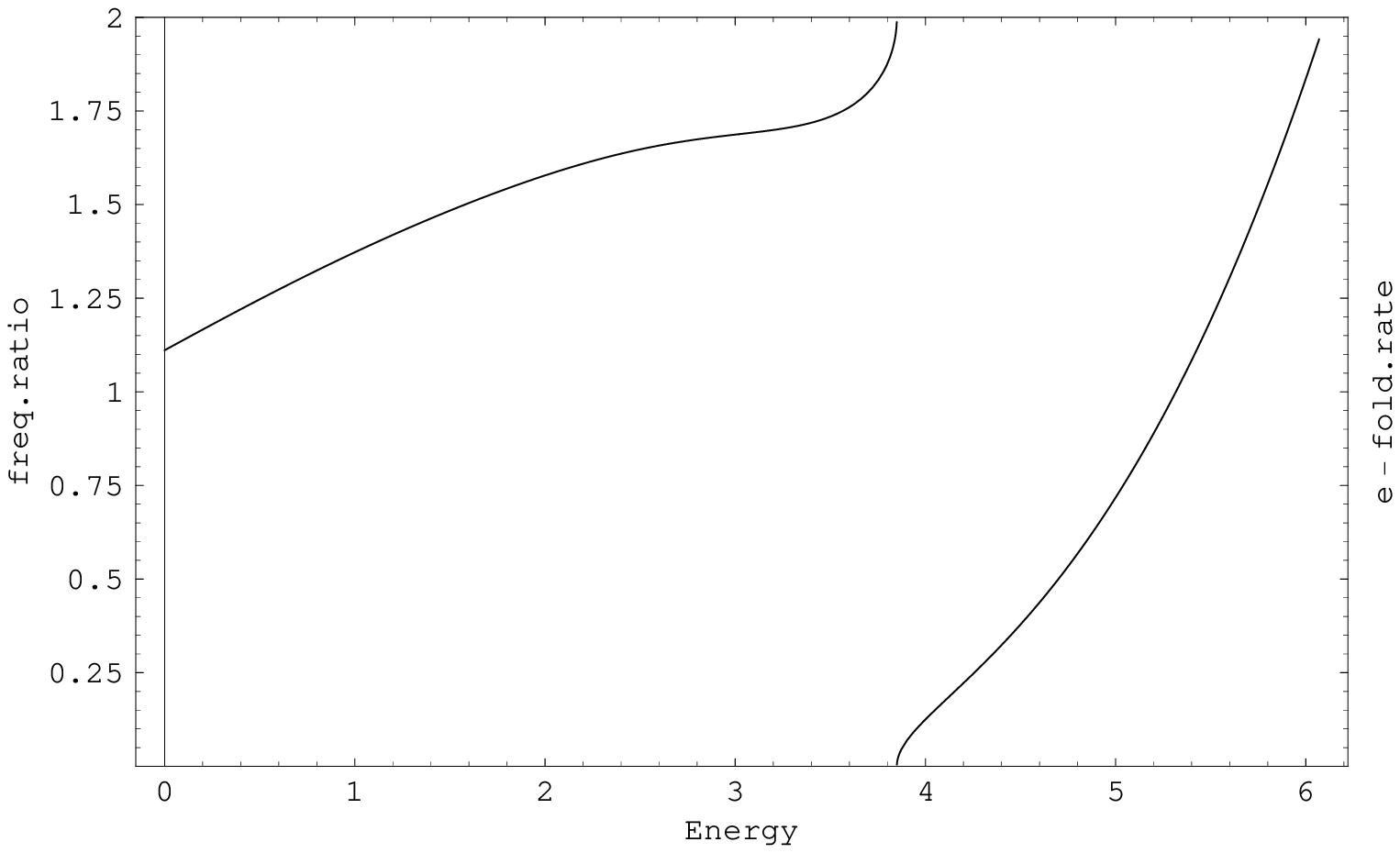}
\caption{Frequency ratio and e-folding rate for the long  {\it x}-axis orbit at $q= 0.9$.}
\label{fig:120_9}
\end{figure}

\begin{figure}
\plotone{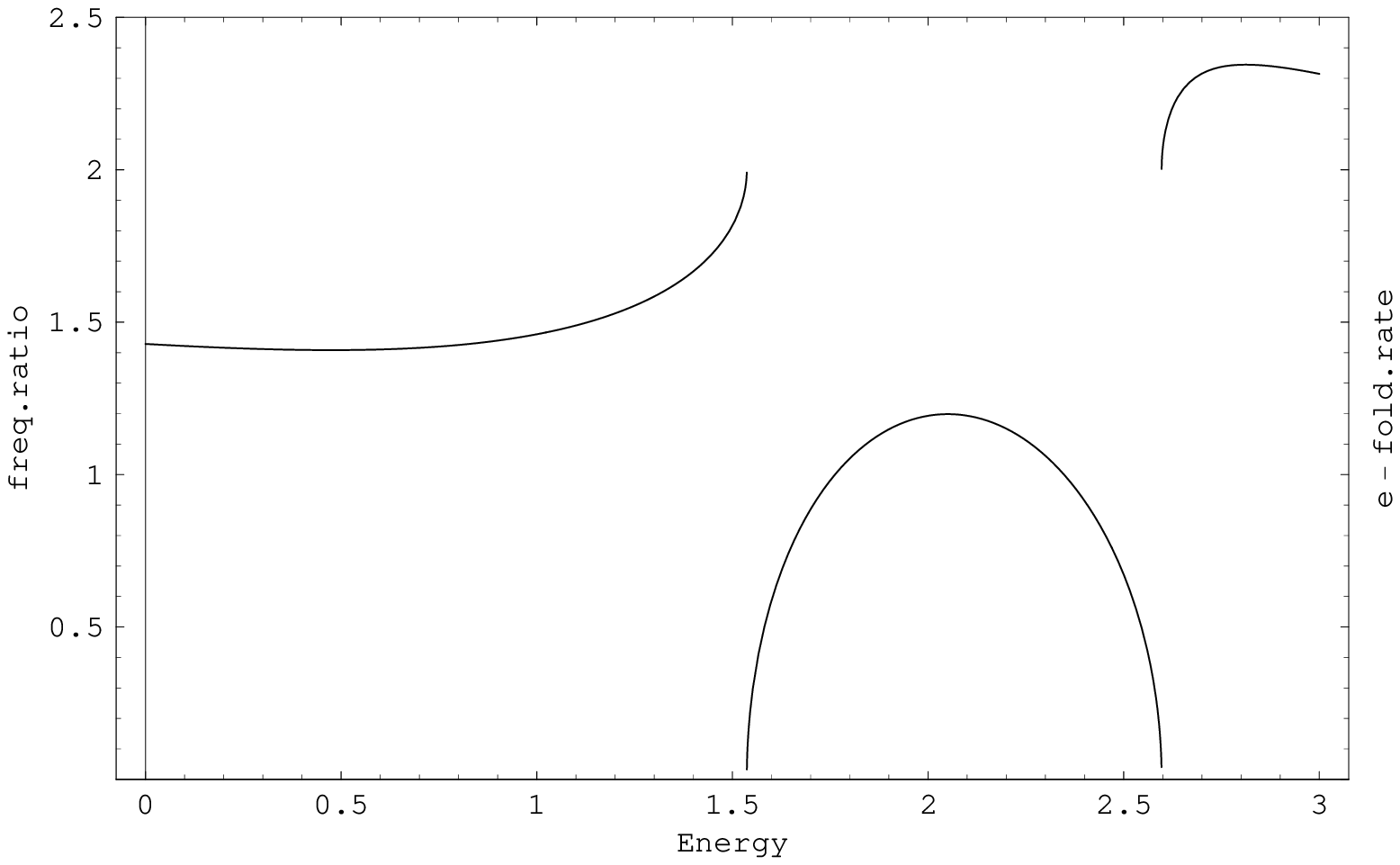}
\caption{Frequency ratio and e-folding rate for the long  {\it x}-axis orbit at $q= 0.7$.}
\label{fig:120_7}
\end{figure}

                       We may also compute the ratio
                       \be\label{freqrat12}
r_1=\frac{\omega_{p}}{\kappa_{1}},\ee 
between the frequency of the perturbation and that of the periodic orbit, where the perturbation frequency is associated to the determinant in the left side of (\ref{d12X}) like in (\ref{pertfreq}) and the e-folding rate as defined in (\ref{efold11}), but with $\kappa_{1}$ in place of $\kappa_{2}$: in this case, at low energy, the ratio has limit $1/q$, since $\omega_{p} \rightarrow 1/q, \kappa_{1} \rightarrow 1$, whereas, at the transition, the ratio has the resonance value 2. In Fig.(\ref{fig:120_9}) the two quantities are plotted for $q=0.9$, in Fig.(\ref{fig:120_7}) they are plotted for $q=0.7$. They can be compared with the corresponding figures in MES (upper panels in their figg.7 and 8) where they have been obtained numerically. 
                
 A demanding test of an analytical approach like that based on normal forms is that of investigating higher-order boxlets: actually the procedure is analogous to that above, but the number of terms in the normal form can be very high. We have limited ourselves to investigating the bifurcation of fish orbits. To this aim we have to compute the 2:3 symmetric normal form which must include terms of degree at least $2\times(2+3)=10$. Once obtained $K^{(2:3)}$, we can express it in adapted resonance coordinates           
           \ba
\psi &=& 6 \theta_1 - 4 \theta_2, \label{psi23}\\
\chi &=& 6 \theta_1 + 4 \theta_2, \label{chi23}\\
J_1 &=& 2{\cal E} + 3 {\cal R}, \label{j123}\\
J_2 &=& 3 {\cal E} - 2 {\cal R}. \label{j223}\ea
and, as before, knowledge of the orbit structure is obtained by investigating the fixed points of the one-dimensional dynamical system associated to the effective Hamiltonian
\be
{\widetilde K}^{(2:3)} = K^{(2:3)}  ({\cal R}, \psi ; {\cal E},q) .\ee
The periodic orbits are identified by the zeros of the system
\ba
&& {\widetilde K}^{(2:3)}_{\cal R} =0, \label{dR23}\\ 
&& {\widetilde K}^{(2:3)}_{\psi} =0. \label{dpsi23}\ea
In particular, (\ref{dpsi23}) is satisfied by $\psi=0$ and the corresponding root of (\ref{dR23})
\be\label{psizero23}
{\cal R} = {\cal R}_{F}({\cal E},q)\ee
determines the ``fish'' orbit. Using this relation and (\ref{j123}) in the existence condition $0\le J_{1F} \le 2{\cal E}$, gives an equation whose roots are related to the bifurcation of the fish orbit from the normal mode. Using as usual the true energy in place of the fictitious ${\cal E}$, we get the fourth degree equation
\be\label{fish}
\sum_{i=0}^{4}c_{i}(q)E^{i}=0,\ee
with coefficients
\footnotesize
 \begin{eqnarray*}
c_{0} &=& -6+9q,\\
c_{1} &=& 3 - \frac{27q}{4},\\
c_{2} &=& -\frac{381}{500} + \frac{54639q}{4000} - \frac{677727q^2}{16000} + \frac{14337q^3}{256} - \frac{12393q^4}{512},\\
c_{3} &=& \frac{637}{250} - \frac{3186q}{125} + \frac{705807q^2}{8000} - \frac{131787 
        q^3}{1024} + \frac{66339 q^4}{1024},\\
c_{4}  &=& -\frac{18689}{12800} + \frac{4279197 q}{256000} + \frac{28267497
q^2}{7168000} - \frac{15582699 q^3}{102400} + \frac{1062755397 q^4}{3276800}
- \frac{1370678193 q^5}{4096000} + \frac{6134535 q^6}{32768} - \frac{5688387
q^7}{131072}.
\end{eqnarray*}
\normalsize
Equation (\ref{fish}) has a complex pair and a real pair of solutions: the smaller positive root can be used for our purpose as a prediction of the bifurcation energy. For $q=0.7$ and $q=0.9$, we get $E_{F}=0.2$ and $E_{F}=1.7$ respectively. The bifurcation energies numerically found in MES are $E_{F}=0.21$ and $E_{F}=2.28$ respectively: clearly, the disagreement between the theoretical and numerical results rapidly grows with detuning.
    
\subsection{Phase space fraction occupied by boxlets}

An analytical estimate of the fraction of phase space occupied by each orbit family can be made with the tools constructed so far. The method is based on the computation of the area on the surface of section associated to the given orbit family. In this respect the estimate is affected by an error due to neglecting a factor depending on the period of individual orbits: this factor in general doesn't allow us to assume simple proportionality between volumes in phase space and areas on the surface of section (Binney, Gerhard \& Hut, 1985). However, in our case, the periods of the families at hand belong to a range small enough to produce a negligible effect. To illustrate the method in its simplest form, we use the lowest order theory applied to loop orbits in Sect. 3 and to bananas in Sect. 4. 

\begin{figure}
   \plotone{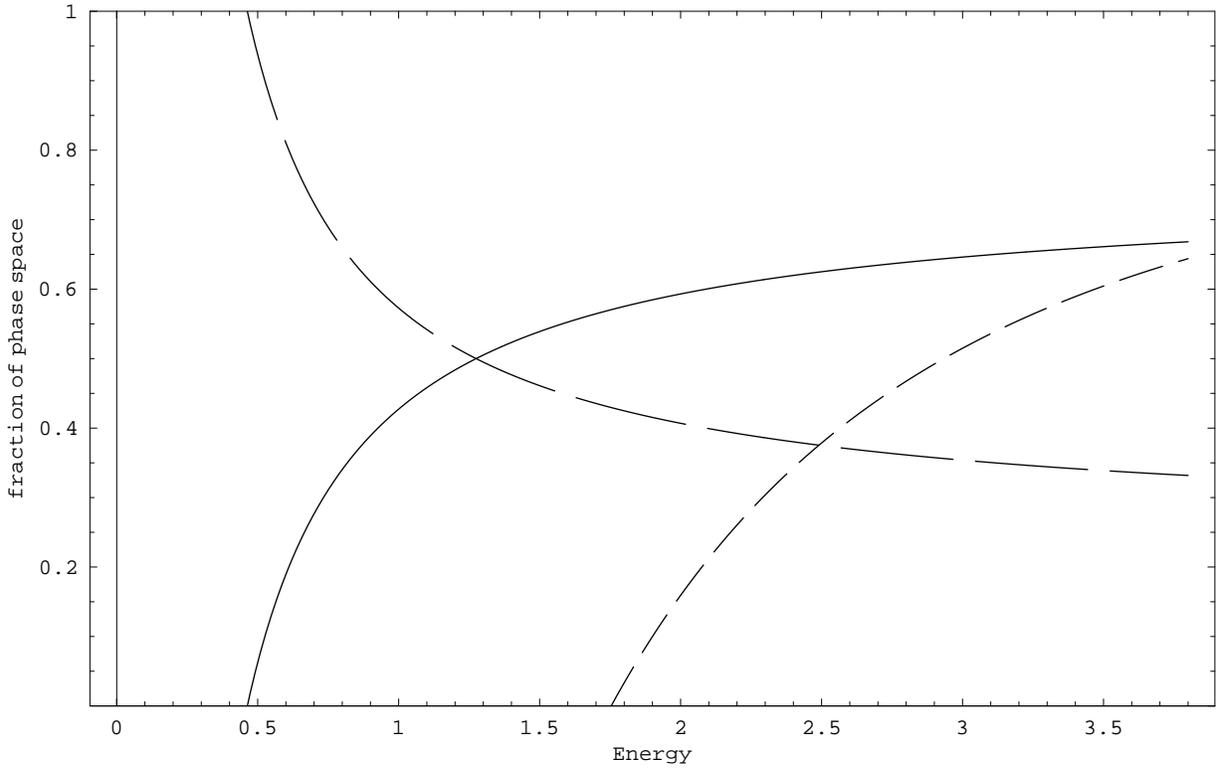}
\caption{Fraction of phase space occupied by low order boxlets at $q= 0.7$: loops (continuous line); boxes (long dashed line); bananas (short dashed line).}
\label{fig:frac0_7}
\end{figure}

\begin{figure}
   \plotone{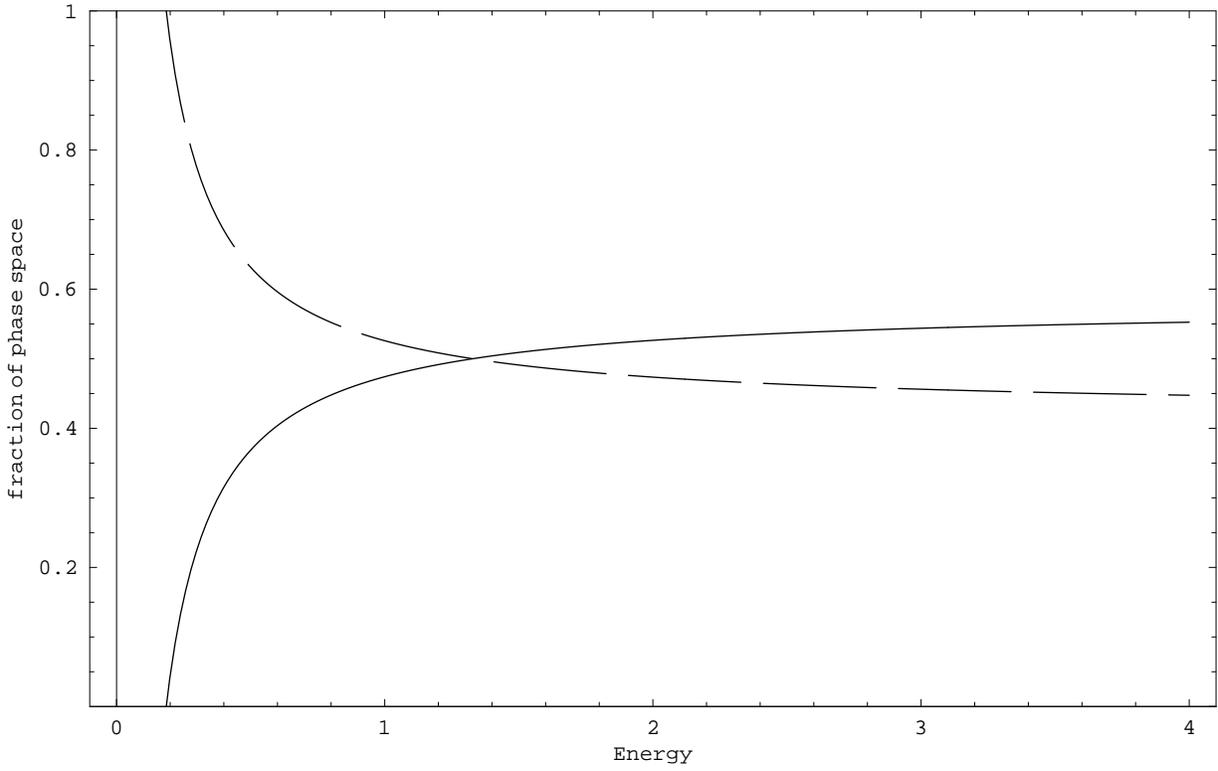}
\caption{Fraction of phase space occupied by low order boxlets at $q= 0.9$: loops (continuous line); boxes (dashed line).}
\label{fig:frac0_9}
\end{figure}

For the loop orbits we have that (\ref{LOOP}) give the values of the actions in terms of $\cal E$ and $q$. As usual, to compare with quantities of the `real' system, we need to convert to the physical energy and this can be done by means of the $K$ on the loop
\be
K^{(1:1)} |_{III} = \frac{q((1+q) {\cal E} - {\cal E}^2 + 2 (q-1)^2)}{3 q^{2}-2q+3} = q E.
\ee
Inverting and expanding, this gives the integral ${\cal E} = {\cal E} (E)$ in terms of the true energy. An important point to remark on, is that the $J's$ appearing in the general expression of the normal form, e.g. $K^{(1:1)}$ of (\ref{HDZM}), are not true actions along a given family. However, the $J's$ in (\ref{LOOP}) {\it are} true actions: in particular, at the bifurcation of the loop from the axial orbit, they are the actions of the orbits asymptotic ({\it homoclinic}) to the $y$-axis orbit, the ``first'' loop and the ``last'' box respectively. Therefore, $J_{1}(E) |_{III}$ is the area inside the homoclinic orbit on the $x-p_x$ surface of section and $J_{2}(E) |_{III}$ is the area inside the homoclinic orbit on the $y-p_y$ surface of section. The ratios
\be\label{frac11}
f_{Loop}=\frac{J_{1}(E) |_{III}}{{\cal E}(E)}, \quad
f_{Box}=\frac{J_{2}(E) |_{III}}{{\cal E}(E)}, \ee
can therefore be used as an approximation of the fraction of phase space occupied by the loops and the boxes respectively. Substituting into (\ref{LOOP}) and dividing by ${\cal E} (E)$, their expressions are
\be\label{fracloop}
f_{Loop} = \frac{2(-3 + 3 q - 5 q^2 + 5 q^3) + E (9 - 9 q + 11 q^2 - 3 q^3)}
                       {(3 - 2 q + 3 q^2)(-2(1-q)^{2}+ E (3 - 2 q + 3 q^2))} \ee
and
\be\label{fracbox}
f_{Box} = \frac{q(10 -10 q + 6 q^2 - 6 q^3 + E (-3 + 11 q - 9 q^2 + 9 q^3))}
          {(3 - 2 q + 3 q^2)(-2(1-q)^{2}+ E (3 - 2 q + 3 q^2))}. \ee
In Figg.(\ref{fig:frac0_7},\ref{fig:frac0_9}) the two quantities are plotted for $q=0.7$ and $q=0.9$ respectively and a comparison can be made with the numerical estimate of $f_{Loop}$ in MES (cfr. their Fig.10) computed by applying an approximate version of the recipe by Binney et al. (1985). We remark that with their approach, MES are not able to compute the fraction pertaining to boxes, since their method does not use families associated to normal modes. 

Concerning banana orbits, we may follow an analogous line of reasoning, obtaining
\be\label{frac12}
f_{Ban}=\frac{J_{1(B)}(E)}{{\cal E}(E)}, \ee
where $J_{1(B)}$, introduced in (\ref{psizero2}), can be interpreted as the area inside the orbit homoclinic to the unstable fixed point on the $x-p_x$ surface of section. In Fig.(\ref{fig:frac0_7}) this quantity is plotted for $q=0.7$ whereas, for $q=0.9$, its application is meaningless because it concerns exceedingly high energies. Comparing with the numerical estimate of $f_{Ban}$ in MES (cfr. the left panel of their Fig.10), we have to take into account that our analytical estimates originate from two independent systems, the two normal forms based on the 1:1 and 1:2 resonances. Therefore, we cannot expect that the occupancies of each family sum up to give the whole phase space. A suitable normalization would be necessary to adjust the proportions and could account for the numerical result. However, since the normalization process contains a certain degree of arbitrariness, we let the quantity as they are and use them only as relative measures. We come again to this point in the next subsection.

\subsection{Surface of section for singular logarithmic potential}

\begin{figure}
  \plotone{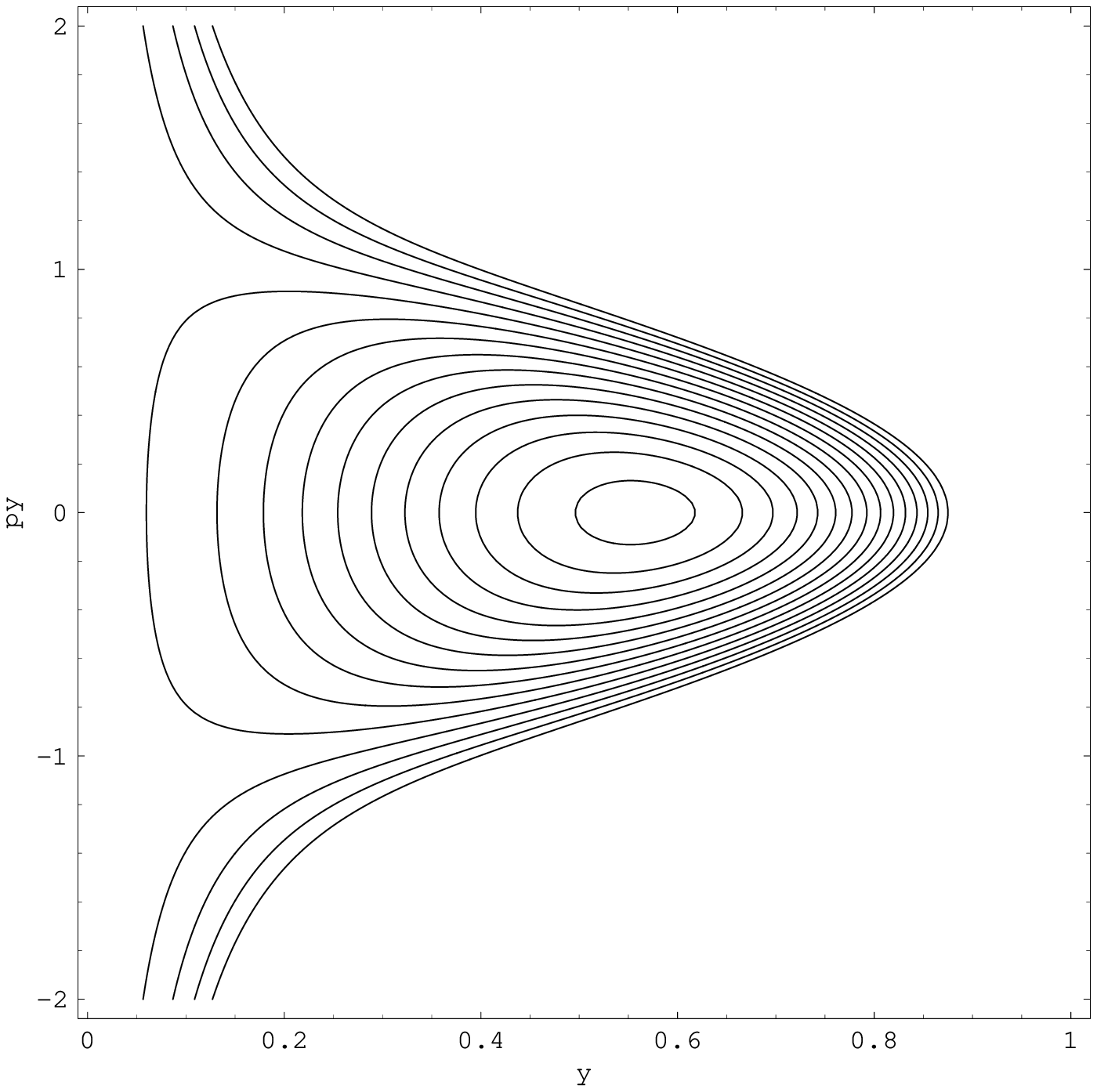}
\caption{Zero-energy $y-p_y$ surface of section of the singular logarithmic potential with $q= 0.7$, computed with the 1:1 resonant normal form.}
\label{fig:psloop}
\end{figure}

\begin{figure}
  \plotone{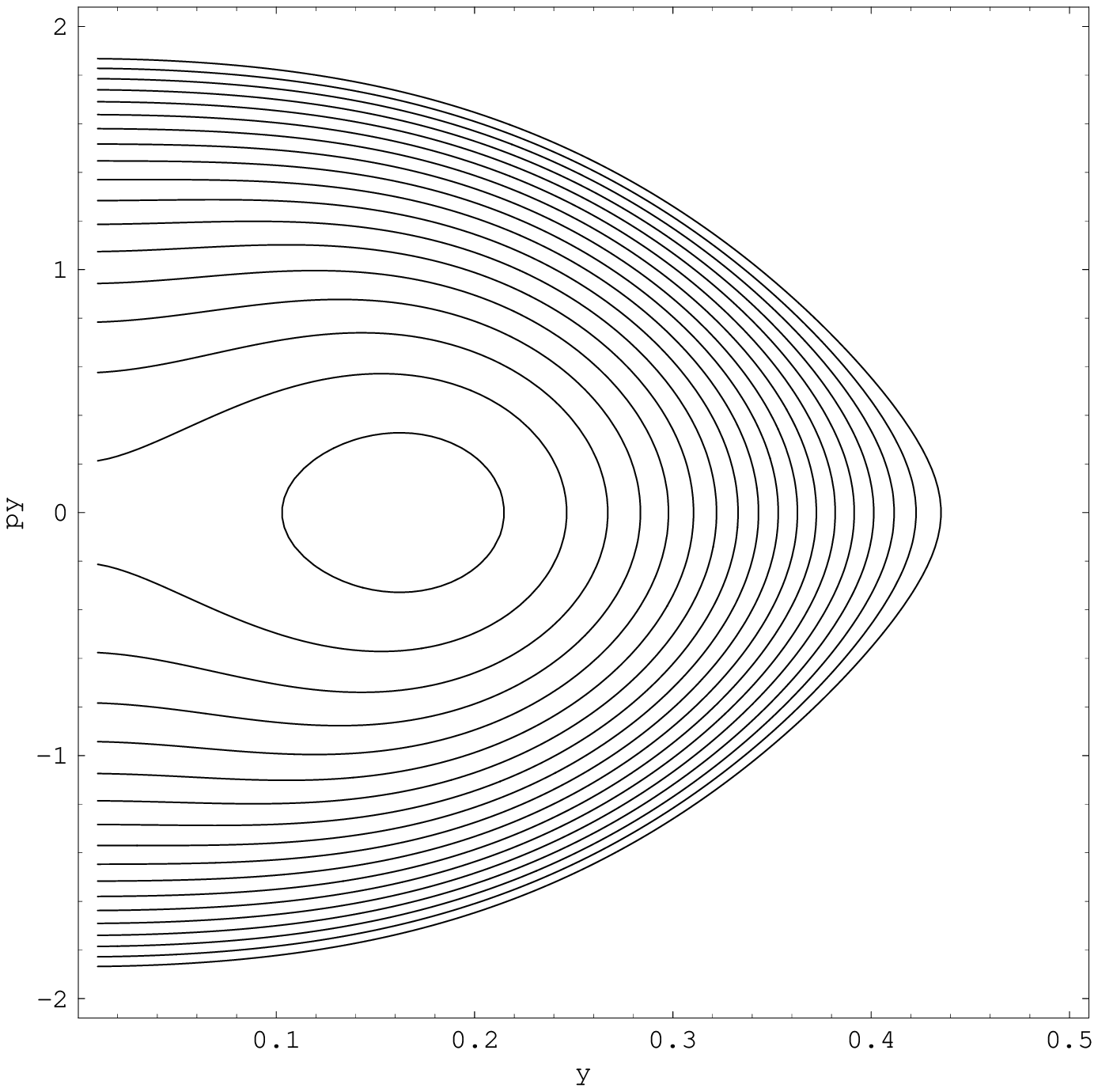}
\caption{Zero-energy $y-p_y$ surface of section of the singular logarithmic potential with $q= 0.7$, computed with the 1:2 resonant normal form.}
\label{fig:psbox}
\end{figure}

It is tempting to try to extract information concerning the scale-free singular limit of the logarithmic potential from our analytical setting based on series expansions. Formally, this operation should be hindered by the lack of a series representation of the singular logarithmic potential. However, we may nonetheless `force' our approximate integrals of motion to play their role in the singular limit too.
If we try our chance by constructing e.g. the $y-p_y$ surface of section with the same procedure as in Subsect. 4.3, using now
\be
\frac{1}{2}(p_x^2+p_y^2) +\frac12 \log(x^2 + y^2 / q^2)=0\ee
to eliminate $p_x$, we get, quite surprisingly, acceptable results. In Fig.(\ref{fig:psloop}) we see the section obtained by using the approximate integral $I^{(1:1)}$ of (\ref{inva2}) for $q=0.7$: it gives the family of loops around the stable periodic orbit at $y \simeq 0.56$ and can be compared with Fig.1 in MES. In Fig.(\ref{fig:psbox}) we see the section obtained by using the approximate integral $I^{(1:2)}$ of (\ref{inva4}), again for $q=0.7$: it gives the family around the stable banana at $y \simeq 0.16$ and boxes around it. A comparison with the same Fig.1 in MES shows that also the banana is located quite well on the section. However, from these considerations emerge the main limitation of the approach based on resonant normal forms: {\it each resonance captures only one feature} of the system under study. For example, comparing our Figg.(\ref{fig:psloop},\ref{fig:psbox}) with a numerically computed section like that in Fig.1 in MES, we see that, in the `true' system, the island around the banana `eats' a substantial part of the island around the closed loop. A reliable description of this phenomenon is not possible in the framework of a {\it single-resonance} theory. This remark is evidently related to the interpretation of the previous results concerning the computation of the fractions in phace space. 


\section{Conclusions}

We have applied the method of resonant detuned normal forms to investigate relevant features of a non-integrable potential of interest in galactic dynamics. The analytical set up consists in constructing a new integrable system (the `normal form') by means of a sequence of canonical transformations with the method of the Lie transform. The algorithm can be put in an efficient and powerful form (Boccaletti \& Pucacco, 1999; Giorgilli, 2002) which can handle quite high-order expansions. We have shown how to exploit resonant normal forms to extract information on several aspects of the dynamics of the original system. In particular, using energy and ellipticity as parameters, we have computed the instability thresholds of axial orbits, bifurcation values of low-order boxlets and phase-space fractions pertaining to the families around them. We have also shown how to infer something about the singular limit of the potential. 

As in any analytical approach, this
method has the virtue of embodying in (more or less) compact formulas simple rules to compute specific properties, giving a general overview of the behavior of the system. In the case in which a non-integrable system has a regular behavior in a large portion of its phase space, a very conservative strategy like the one adopted in this work provides sufficient qualitative and quantitative agreement with other more accurate but less general approaches. In our view, the most relevant limitation of this approach, common to all perturbation methods, comes from the intrinsic structure of the single-resonance normal form. The usual feeling about the problems posed by non-integrable dynamics is in general grounded on trying to cope with the interaction of (several) resonances. Each normal form is instead able to correctly describe only one resonance at the time. However, we remark that the regular dynamics of a non-integrable system can be imagined as a superposition of very weakly interacting resonances. If we are not interested in the thin stochastic layers in the regular regime, each portion of phase space associated with a given resonance has a fairly good alias in the corresponding normal form. An important subject of investigation would therefore be that of including weak interactions in a sort of higher order perturbation theory. For the time being, there are two natural lines of developement of this work: 1. to extend the analysis to cuspy potentials and/or central `black holes'; 2. to apply this normalization algorithm to three degrees of freedom systems.

\clearpage

\end{document}